\documentstyle[12pt]{article}
\newcommand{\la}[1]{\label{#1}}

\newcommand{\G}{{\cal G}}

\newcommand{\cc}{{\bf c}}
\newcommand{\bb}{{\bf b}}
\newcommand{\uu}{{\bf u}}
\newcommand{\vv}{{\bf v}}

\newcommand{\X}{{\cal X}}

\newcommand{\non}{\nonumber}
\newcommand{\be}{\begin{equation}}
\newcommand{\ee}{\end{equation}}
\newcommand{\ba}{\begin{eqnarray}}
\newcommand{\ea}{\end{eqnarray}}
\newcommand{\bastar}{\begin{eqnarray*}}
\newcommand{\eastar}{\end{eqnarray*}}
\newcommand{\half}{{1 \over 2}}
\begin{document}
\begin{titlepage}

\begin{flushright}
UU-ITP-06/96 \\
\end{flushright}

\vskip 0.5truecm

\begin{center}
{ \bf \Large \bf Aspects of mass gap, confinement and \\
                 
\vskip 0.2cm
                N=2 structure in 4-D Yang-Mills theory \\
}
\end{center}

\vskip 0.7cm

\begin{center}
{\bf Jussi Kalkkinen$^{*}$ and Antti 
J. Niemi$^{*\dagger}$ } \\

\vskip 0.6cm

{\it Department of Theoretical Physics,
Uppsala University \\
P.O. Box 803, S-75108, Uppsala, Sweden  \\
}
\end{center}

\vskip 1.5cm
\rm
\noindent
We introduce new variables in
four dimensional SU(N)
Yang-Mills theory. These variables emerge when
we sum the path integral
over classical solutions and represent
the summation as an
integral over appropriate degrees of freedom.
In this way we get an effective field
theory with SU(N)$\times$SU(N) gauge symmetry.
In the instanton approximation our effective
theory has in addition a N=2 supersymmetry, and 
when we sum over all possible solutions 
we find a Parisi-Sourlas supersymmetry. These extra
symmetries can then be broken explicitly  by a
SU(N) invariant and power counting renormalizable 
mass term. Our results suggest that the confinement 
mechanism which has been  recently identified 
in the N=2 supersymmetric 
Yang-Mills theory might also help to understand
color confinement in ordinary, pure 
Yang-Mills theory. In particular, there appears to 
be an intimate relationship 
between the N=2 supersymmetry
approach to confinement and the
Parisi-Sourlas dimensional reduction. 
\vfill

\begin{flushleft}
\rule{5.1 in}{.007 in} \\
$^{\dagger}$ \small Supported by G{\"o}ran Gustafsson
Foundation for Science and Medicine \\
\hskip 0.3cm and by NFR Grant F-AA/FU 06821-308
\\ \vskip 0.2cm
$^{*}$  \small E-mail: \scriptsize
\bf JUSSI.KALKKINEN@TEORFYS.UU.SE $~~$ 
ANTTI.NIEMI@TEORFYS.UU.SE  \\
\end{flushleft}

\end{titlepage}
\vfill\eject
\baselineskip 0.65cm

\noindent
\section{Introduction}

Recently, there has been impressive progress 
in understanding confinement in 
supersymmetric Yang-Mills theories \cite{nati}.
In particular, 
the N=2 supersymmetric theory with its
supersymmetry properly broken to N=1, 
appears to materialize
the qualitative picture of 
confinement introduced in 
\cite{thooft}. However, despite remarkable
success it remains a challenge
to extend these results to standard QCD where
we still lack a convincing explanation why color
and quarks confine.

Intimately related to the problem of
color confinement is the origin 
of a mass gap in Yang-Mills theory. 
In the case of N=2 supersymmetric Yang-Mills
theory a mass gap is introduced explicitly, by
adding a mass term to the scalar multiplet \cite{nati}.
But in the case of ordinary Yang-Mills theory 
elementary scalar 
fields are absent. Instead, a mass gap
is supposed to have a dynamic origin in the infrared
divergence structure of the theory. However, at
the moment we still lack a convincing explanation
how the mass gap actually appears.

In the present paper we are interested in
the problem of color confinement and 
the related emergence of a mass gap
in ordinary, pure Yang-Mills theory. For this
we shall investigate how the results derived in the
supersymmetric context can be extended to ordinary
Yang-Mills theory. In particular, we inquire 
whether a proper variant of the N=2 structure 
can be identified, and whether a gauge invariant 
and renormalizable mass scale can be explicitly
introduced. 

Since the number of degrees of freedom
in ordinary Yang-Mills theory is insufficient for
identifying structures such as the N=2 supersymmetry,
we need to introduce additional
variables. For this we investigate the nonperturbative
structure of Yang-Mills path 
integral in the background field formalism 
\cite{bff}. We first separate 
the classical background fields 
from their
quantum fluctuations, and explicitly sum the path 
integral over these classical
backgrounds. We then represent this summation as 
a path integral over a set of auxiliary variables. 
This promotes the classical background 
configurations into {\it off-shell} degrees of freedom,
and introduces a large number 
of new variables. These
variables
determine a topological quantum
field theory that describes the space of classical 
solutions of the original Yang-Mills theory.  
We apply standard BRST 
arguments to extend this topological field theory so 
that it involves a number of parameters, with BRST
symmetry ensuring that the theory is
independent of these parameters. For a
particular value of these parameters we recover 
explicitly the original summation over classical 
Yang-Mills fields in the background field formalism. 
But for other values of these parameters our 
effective field theory admits other interpretations.

We shall be interested in two a priori different sets 
of classical field configurations, the (anti)selfdual
ones and those that describe all possible solutions to
the classical Yang-Mills equations.

We first consider the (anti)selfdual configurations.
For a finite Euclidean action these are the instantons,
and the pertinent topological field theory is 
the one introduced
by Witten to describe Donaldson's invariants
\cite{witten}. 
When we sum our
Yang-Mills path integral over
all (anti)selfdual 
configurations in the background field
formalism, we then obtain 
an effective field theory that
describes a coupling between the original 
Yang-Mills quantum degrees of freedom and
Witten's topological Yang-Mills theory.
Furthermore, when we employ unitarity arguments 
to introduce a twist in the Lorentz transformations 
we can identify our topological sector as the 
minimal N=2 supersymmetric Yang-Mills theory.

This appearance of a N=2 structure in
ordinary Yang-Mills theory is
certainly most interesting. We view it
as an indication that the results derived in 
\cite{nati} could be extended to explain color 
confinement in ordinary, pure Yang-Mills theory. 
However, we first  need to understand 
how a mass gap appears. For this we observe
that besides 
the N=2 supersymmetry, the topological 
sector of our theory also admits an
additional SU(N) gauge invariance implying that
our effective theory actually
has an extended SU(N)$\times$SU(N) gauge invariance.
Both of these additional 
symmetries may then be explicitly
broken to the original SU(N) 
symmetry of the ordinary Yang-Mills
theory. This opens the possibility
that we can introduce 
additional terms to our effective action,
provided these 
additional terms are consistent
with the SU(N) gauge invariance 
of our original Yang-Mills theory. Such additional
terms might then have interesting physical consequences
to the original Yang-Mills theory.

One possibility to break part of 
these extra symmetries
is, that following \cite{nati} we
add an explicit mass term to the scalar superfield.
This breaks the N=2 supersymmetry in the 
topological sector explicitly 
into a N=1 supersymmetry.
Since this mass term is
renormalizable, in the present context it 
might  also provide an alternative 
to the standard Higgs 
mechanism for constructing a
renormalizable theory of massive gauge vectors.
However, a disadvantage of 
such a scalar superfield mass term
is that it does not break the SU(N)$\times$SU(N) 
gauge invariance into the 
physical SU(N). For this reason
we propose an alternative, which arises when 
we add a mass term to the field that describes 
quantum fluctuations around the original classical 
background. This mass term explicitly breaks both the
N=2 supersymmetry and the SU(N)$\times$SU(N) gauge
symmetry into the physical SU(N) gauge symmetry.
We argue that this mass term is also 
consistent with unitarity and power counting
renormalizability, suggesting that it might
have some relevance to the expected generation of a 
mass gap in ordinary quantum Yang-Mills theory.
Furthermore, this mass term might 
also provide an alternative to the
Higgs mechanism for constructing massive gauge
vectors. 

We then proceed to investigate 
the effective theory that appears
when we sum over all possible solutions to the Yang-Mills
equation. In addition of instantons, now  
the pertinent topological quantum field theory also
describes classical solutions that do not necessarily
satisfy the first-order (anti)selfduality equation. As a
consequence, instead of the N=2 supersymmetry we now find
the Parisi-Sourlas supersymmetry \cite{parisi}
in the topological sector. 
Furthermore, as in the (anti)selfdual case we
find an additional, independent SU(N) gauge invariance
in this topological sector, and 
again both of these additional 
symmetries can be broken explicitly into the
physical SU(N) gauge symmetry. For this, we again introduce
a mass term to the field that
describes quantum fluctuations around the
classical Yang-Mills background.

The emergence of a Parisi-Sourlas supersymmetry
is particularly interesting, since it suggests that
ordinary Yang-Mills theory admits a 
subsector which exhibits the D=4
$\to$ D=2 Parisi-Sourlas dimensional
reduction \cite{parisi}. 
Indeed, previously it has been conjectured that 
such a subsector should
exist \cite{kopis} and it should dominate the Yang-Mills
theory in the infrared limit and
explain color confinement:
When planar Wilson loops
are restricted to this subsector they exhibit an
area law as a consequence of the D=4 $\to$ D=2
dimensional reduction \cite{kupi}, \cite{maillet}. 
The present construction reveals that
this structure actually appears in ordinary Yang-Mills
theory. We verify that at large spatial distances
the propagator for the Parisi-Sourlas 
gauge field behaves like ${\cal O}(p^{-4})$
consistent with a linearly growing potential,
and provide arguments which support that
this behaviour indeed dominates in the infrared limit. We
also explain how the ensuing string tension emerges from 
a cohomological construction. 

Finally, we confirm that the summation over 
(anti)selfdual configurations provides a good 
approximation to the summation over all possible 
classical solutions. This suggests that the approach 
to confinement in supersymmetric theories developed
in \cite{nati} can also be adopted to explain color
confinement in ordinary Yang-Mills theory. In particular,
our results suggest that the N=2 supersymmetry
approach to confinement
is intimately connected to the picture of color 
confinement due to randomly distributed color-electric and
color-magnetic fields with the ensuing Parisi-Sourlas
dimensional reduction as
developed in \cite{kopis}, \cite{kupi}, \cite{maillet}.

\vskip 0.3cm
In the next section we shall shortly review the
background field formalism. We also describe how it can
be applied to implement a summation
over classical field configurations. 
In sections 3. and 4. we 
consider the approximation that arises when
we sum over all (anti)selfdual connections.
In section 5. we explain how a 
gauge invariant mass term
can be introduced, and argue that this mass term also 
preserves power-counting renormalizability.
In section 6. we investigate unitarity of our theory.
We find a manifestly unitary representation by
re-interpreting the action of Lorentz group on 
our fields, and
conclude that ordinary Yang-Mills theory actually
{\it contains} the N=2
supersymmetric Yang-Mills theory. 
In sections 7. and 8. we
proceed to the general case, obtained when
we sum over all possible classical solutions.
Now we find an effective action with 
Parisi-Sourlas supersymmetry. This suggests that
ordinary Yang-Mills theory admits
a nonperturbative sector that 
exhibits the D=4 $\to$ D=2 
Parisi-Sourlas dimensional
reduction. We argue that this sector dominates in the
infrared limit, indicating that 
color confinement could be
a consequence of an effective dimensional reduction.
Finally, in section 9. we verify that
the results obtained in the (anti)selfdual approximation
are consistent with those obtained when we sum over
all possible solutions. This means that the N=2 picture
indeed provides a good desription 
of the exact theory. It
suggests that the recent approach to 
confinement in the N=2
supersymmetric theory may
be directly relevant for describing color confinement 
also in ordinary Yang-Mills theory.

\vskip 0.3cm

\noindent
\section{Background field quantization}

\vskip 0.2cm

We shall consider a
$SU(N)$ Yang-Mills theory with
gauge field $C_\mu^a$ and field strength
\[
F_{\mu\nu} ~=~ [D_\mu , D_\nu ] ~=~
\partial_\mu C_\nu - \partial_\nu C_\mu + 
[C_\mu , C_\nu]
\]
We normalize 
the Lie algebra generators $T^a$ so that 
the classical action is
\be
S_{YM}(C) ~=~ \int  \frac{1}{4}
F^a_{\mu\nu} F^a_{\mu\nu} ~\equiv~
\int \frac{1}{4} F_{\mu\nu}^2
\la{YM0}
\ee
We shall be interested in the (Euclidean) partition
function
\be
Z ~=~ \int [dC] \exp \{ - S_{YM}(C) \}
\la{SYM}
\ee
in the background field formalism. For this
we represent 
the gauge field $C_\mu$ as a linear combination
\be
C_\mu ~=~ A_\mu ~+~ Q_\mu
\la{new1}
\ee
where we select 
$A_\mu$ to satisfy the classical 
equation of motion, 
\be
D_\mu F_{\mu\nu} (A) ~=~ \partial_\mu F_{\mu\nu} +
[ A_\mu , F_{\mu \nu} ] ~=~ 0
\la{eom}
\ee
The field $Q_\mu$ then describes perturbative quantum
fluctuations around the classical field
$A_\mu$. This means that the classical 
field $A_\mu$ satisfies nontrivial
boundary conditions, while the boundary conditions
for the fluctuation field $Q_\mu$ are trivial.
For example in the case
of an instanton the second Chern class 
\[
{\rm Ch}_2(A) \ = \ 
\int F_{\mu\nu} {\tilde F}_{\mu\nu}(A)
\]
for $A_\mu$ is nontrivial and coincides with the
second Chern class for $C_\mu$.

If we define 
\be
G_{\mu\nu} ~=~ D_\mu Q_\nu - D_\nu Q_\mu
\la{G}
\ee
where the covariant derivative is {\it w.r.t.}
the classical field $A_\mu$,
we can write the action (\ref{YM0})
as
\[
- S_{YM}(A+Q) ~=~ - \frac{1}{4}  F_{\mu\nu}^2 \ - 
\ Q_\mu D_\nu
F_{\mu\nu} \ - \ 
\half F_{\mu\nu} [Q_\mu , Q_\nu ]
\]
\be
- \ \frac{1}{4} G_{\mu\nu}^2 \ - \ \half  G_{\mu\nu}
[Q_\mu , Q_\nu ] \ - \frac{1}{4} \  [Q_\mu , Q_\nu ]^2
\
\la{ym}
\ee
Since $A_\mu$ solves the Yang-Mills
equation (\ref{eom}),
the $Q$-linear term $ Q_\mu D_\nu F_{\mu\nu}$
in (\ref{ym}) actually vanishes. However, eventually
we shall promote $A_\mu$ to an off-shell field 
so that it will cease to be
constrained by (\ref{eom}). In that case the
$Q$-linear term does not vanish, and in anticipation
of this we have included it also here.

Notice that there is certain latitude in defining
the gauge transformations of the fields that
appear on the {\it r.h.s.} of (\ref{new1}).
Here we have selected
the gauge transformation of the classical field
$A_\mu$ to coincide with the standard 
gauge transformation of a $SU(N)$ gauge field
\be
A_\mu ~ \to ~ U A_\mu U^{-1} \ + \ U
\partial_\mu U^{-1}
\la{gaugecl}
\ee
so that the fluctuation field $Q_\mu$ gauge
transforms homogeneously, like a Higgs field
\be
Q_\mu ~\to~ U Q_\mu U^{-1}
\la{gaugefl}
\ee
This implies that each term in (\ref{ym}) is
separately gauge invariant.

In terms of these background
variables the path integral 
(\ref{SYM}) becomes
\[
Z_{YM} ~=~ \sum\limits_{A_\mu} \int [dQ]
\exp \{  \int -
\frac{1}{4} F_{\mu\nu}^2 \ - \  
Q_\mu D_\nu F_{\mu\nu} \
- \ \half F_{\mu\nu} [Q_\mu , Q_\nu ]
\]
\be
- \ \frac{1}{4}  G_{\mu\nu}^2 \ - \ 
\half G_{\mu\nu}
[Q_\mu , Q_\nu ] \ - \frac{1}{4} \  
[Q_\mu , Q_\nu ]^2 \ \}
\la{path1}
\ee
The summation extends over all
solutions $A_\mu$ 
of the Yang-Mills equation of motion, and we remind 
that the integral over $Q_\mu$ is
subject to trivial boundary conditions. 
Notice that since the
moduli space of the classical 
field $A_\mu$ is generically 
nontrivial ({\it e.g.} for a $k$-instanton
it is a $8k-3$ dimensional manifold), 
the summation over $A_\mu$ should 
actually be viewed as
an integration over the relevant moduli.

Heuristically, we can 
represent the summation over $A_\mu$ as
\be
\sum\limits_{A_\mu} ~=~ \int [dA] 
\delta( D_\mu F_{\mu\nu} ) \left| \ \det ||
{ \delta D_{\mu} F_{\mu\nu} \over 
\delta A_\rho } || \ \right| ~\approx~
\sum\limits_{A_\mu}
\left| sign \det || { \delta DF \over 
\delta A } || \right|
\la{GBCA}
\ee
but unfortunately, due to the absolute values 
it is usually quite difficult to
implement (\ref{GBCA}) in (\ref{path1}). 
For this reason,
we use the common practice and
replace the summation over $A_\mu$ by
the more manageable
\be
\sum\limits_{A_\mu} ~\to~ \int [dA] 
\delta( D_\mu F_{\mu\nu} )  \det ||
{ \delta D_{\mu} F_{\mu\nu} \over 
\delta A_\rho } || ~\approx~ \sum\limits_{A_\mu}
sign \det || { \delta DF \over \delta A } ||
\la{GBC1}
\ee
and {\it approximate} (\ref{path1}) by
\[
Z_{YM} ~\approx~ \int [dA] [dQ] 
\delta( D_\mu F_{\mu\nu} ) \det ||
{ \delta D_{\mu} F_{\mu\nu} 
\over \delta A_\rho } || \ 
\exp \{  \int - \frac{1}{4}  F_{\mu\nu}^2 \ - \  
Q_\mu D_\nu F_{\mu\nu} \ - 
\  \half F_{\mu\nu} [Q_\mu , Q_\nu ]
\]
\be
- \ \frac{1}{4}  G_{\mu\nu}^2 
\ - \  \half G_{\mu\nu}
[Q_\mu , Q_\nu ] \ - \frac{1}{4} \  
[Q_\mu , Q_\nu ]^2 \ \}
\la{path2}
\ee

\be
\approx \ \sum\limits_{A_\mu} sign \det || { 
\delta D F \over \delta A } || \left\{ \int [dQ]
\exp \{ - \int \frac{1}{4} 
F_{\mu\nu}^2 (A+Q) \} \right\}
\la{pathtwo}
\ee
In order to interpret (\ref{path2}), 
(\ref{pathtwo}) mathematically,
we resort to a finite dimensional 
analogue and
view the Yang-Mills 
action $S_{YM}(A+Q)$ in (\ref{pathtwo})
as an infinite dimensional
counterpart of a (nondegenerate) 
Morse function $H(x)$.
A sum such as (\ref{GBCA}) counts 
the number of its critical
points, and is bounded from below by the
sum of Betti numbers $B_n$ 
of the underlying manifold
(in our case the gauge orbit space
${\cal A}/{\cal G}$)
\[
\sum\limits_{dH=0} 1 \ \geq \ \sum\limits_{n} B_n
\]
On the other hand, a sum such as the {\it 
r.h.s.} of (\ref{GBC1}) is independent of the Morse
function and according to the 
Poincar\'e-Hopf theorem \cite{morse} it
coincides with the Euler character $\X$ of the 
manifold
\be
\sum\limits_{dH=0} sign \det || { \partial^2 H \over 
\partial x_a \partial x_b } || ~=~ \sum_n (-)^n B_n \
\equiv \ \X
\la{eulerchar}
\ee
If $H$ is a perfect Morse function these two 
quantities coincide, but for a 
general Morse function
they are different since in general the 
fluctuation matrix $ \partial_{ab}H$ admits an 
odd number of zeromodes for some of the
critical points $x_a$.

For a compact finite dimensional Riemannian manifold
the Euler character (\ref{eulerchar})
coincides with
the partition function of the de Rham supersymmetric
quantum mechanics \cite{wit2},
\cite{blau}. This partition function
can be evaluated exactly, {\it e.g.} by localizing the
corresponding path integral to
the Euler class of the manifold.
In this way we obtain
the standard
relation between the Poincar\'e-Hopf
and Gauss-Bonnet-Chern theorems.

On the other hand, the summation that appears 
in (\ref{path2}) is an 
infinite dimensional generalization of a sum
of the form
\[
\sum\limits_{dH=0} sign \det || { \partial^2 H \over 
\partial x_a \partial x_b } || \ \exp \{ - T {\cal H} \}
\]
where ${\cal H}$ corresponds to the $Q$-integral
in (\ref{pathtwo}). When $H$ and ${\cal H}$ coincide, 
we obtain a quantity
that appears in an equivariant version of the
Poincar\'e-Hopf theorem \cite{palo}.
There is also an equivariant version of the 
Gauss-Bonnet-Chern theorem, and as in 
conventional Morse theory one can derive a relation
between these two theorems using 
an equivariant version of the de Rham supersymmetric 
quantum mechanics \cite{palo}. The pertinent 
path integral is intimately related to a standard
Hamiltonian path integral,
with the Morse function $H$ interpreted
as the Hamiltonian function.
Indeed, this
interrelationship between equivariant Morse
theory and standard 
Hamiltonian path integrals can be
utilized to evaluate certain
Hamiltonian partition functions
exactly, using localization 
techniques \cite{dh}. This
leads to the Duistermaat-Heckman integration \cite{bgv}
formula and its quantum 
mechanical generalizations \cite{palo}, \cite{dh}. 

The present mathematical interpretations 
suggest that (\ref{path2}) may
in fact provide a relatively good approximation
of the original partition function (\ref{path1}). 
Proceeding with (\ref{path2}) we at least
obtain a reliable lower-bound estimate of the 
partition function. In the following we shall
investigate (\ref{path2}), using the insight
provided by these mathematical analogues.

\vskip 0.3cm

The Yang-Mills equation (\ref{eom}) admits 
various different kinds of solutions, and the most
interesting ones are the instantons. 
For an instanton the fluctuation 
matrix in (\ref{GBCA}), (\ref{path2}) 
admits only non-negative eigenvalues. 
Consequently in an instanton approximation
the absolute signs in (\ref{GBCA}) are 
not relevant, implying that (\ref{path2}) describes 
adequately the pertinent contribution
to the original path integral (provided 
we properly account for
the zero modes in 
the fluctuation matrix). This
is intimately related to the fact, that for instantons
the Yang-Mills 
equation simplifies
to the first-order (anti)selfduality equations
\be
F^{\pm} ~=~ F_{\mu\nu} \ \pm \ \half 
\epsilon_{\mu\nu\rho\sigma}
F_{\rho\sigma} ~=~ 0
\la{ainst}
\ee
where the $+$ refers to anti-selfdual configurations
and $-$ to selfdual ones.

Formal Morse theory 
arguments suggest that the Poincar\'e-Hopf 
representation of
the Euler character $\X$ on ${\cal A}/{\cal G}$ 
should be independent of the
Morse function. This means, that we should be able to
utilize the (anti)selfduality
equations (\ref{ainst}) to introduce the following
alternative realizations of the Euler character,
\[
\X ( {\cal A}/{\cal G} ) ~=~
\sum\limits_{DF=0} sign \det || { \delta DF 
\over \delta A}|| ~=~ 
\sum\limits_{ F^{-} = 0 }
sign \det  ||
{ \delta  F^-  \over 
\delta A } || ~
=~ \sum\limits_{ F^{+} = 0 }
sign \det  ||
{ \delta  F^+  \over 
\delta A } || 
\]
\be
\approx ~ 
\int [dA] \delta( F^{\pm} ) \det ||
{ \delta  F^{\pm} \over 
\delta A } || 
\la{GBC2A}
\ee
Furthermore, since the (anti)selfdual 
configurations solve the
original Yang-Mills equation, this suggests
that we might also obtain a reliable 
approximation to (\ref{path1}) if instead 
of (\ref{path2}),
(\ref{pathtwo}) we use
\[
Z_{YM} ~\approx~
\sum\limits_{F^{\pm} = 0} sign \det || { 
\delta F^{\pm} \over \delta A } || \left\{
\int [dQ] \exp \{ \int - 
\frac{1}{4} F_{\mu\nu}^2 (A+Q) \} \right\}
\]
\[
\approx \ \int [dA] [dQ] 
\delta( F^{\pm} ) \det ||
{ \delta F^{\pm} 
\over \delta A } || \ 
\exp \{  \int
- \frac{1}{4}  F_{\mu\nu}^2 \  - \  
Q_\mu D_\nu F_{\mu\nu}
\ - \  \half F_{\mu\nu} [Q_\mu , Q_\nu ]
\]
\be
- \ \frac{1}{4}  G_{\mu\nu}^2 \ - \ \half G_{\mu\nu}
[Q_\mu , Q_\nu ] \ - \frac{1}{4} \  
[Q_\mu , Q_\nu ]^2 \ \}
\la{path3}
\ee
where the zeromodes can again be
accounted for {\it e.g.} 
by introducing collective coordinates.

Notice that since the (anti)selfdual 
fluctuation matrices that appear in (\ref{path3})
are first-order operators,
these matrices admit an infinite number 
of negative eigenvalues. In particular this
means that (\ref{path3})
is {\it a priori} quite different from the usual
semiclassical approximation, where one evaluates
the fluctuations around {\it both} 
instantons and anti-instantons using the path
integral (\ref{path2}). There, the fluctuation
matrix is the original
Yang-Mills one and for (anti)selfdual configurations
it is positive semidefinite.

Finally, we inquire how (\ref{path2})
should be corrected so that
instead of approximating the original 
path integral (\ref{path1}),
we get exact results. For this, we need to
account for the
absolute signs in the fluctuation determinants. 

Formally, the sign of the fluctuation determinant
in (\ref{GBCA}) is
\be
sign \det ||  { \delta DF \over 
\delta A } || ~=~
\exp\{ i \pi \sum\limits_{ \lambda_n < 0 } 1 \}
\la{phases1}
\ee
where $\lambda_n$ are eigenvalues of the fluctuation
matrix,
\[
{\delta D F \over 
\delta A } \psi_n ~=~ \lambda_n \psi_n
\]
We write
\[
\sum\limits_{\lambda_n < 0 } 1 ~=~ 
\half \sum\limits_{\lambda_n} 1
\ - \ \half \sum\limits_{\lambda_n} sign( \lambda_n )
\]
\[
\ \ = ~ \half \zeta_{YM} \ - \ \half \eta_{YM}
\]
where $\zeta_{YM}$ is the $\zeta$-function 
of the fluctuation
matrix
\[
\zeta_{YM}(s) \ = \ \sum\limits_{\lambda_n} 
| \lambda_n |^{-s}
\]
and $\eta_{YM}$ is its $\eta$-invariant,
\[
\eta_{YM}(s) \ = \ \sum\limits_{\lambda_n}
sign(\lambda_n) | \lambda_n |^{-s}
\]
both evaluated at $s=0$ and at the background 
configuration $A_\mu$.
Consequently the phase is
\be
sign \det || { \delta DF \over \delta A } || ~=~
\exp \{ \frac{i}{2} \pi ( \zeta_{YM} - \eta_{YM} ) \}
\la{phases2}
\ee
This implies that we obtain (\ref{path1}) if 
we improve (\ref{path2}) to
\[
Z_{YM} \ = \ \int [dA] [dQ] \delta ( D F ) \det ||
{\delta DF \over \delta A } || \exp \{ \int - \frac{1}{4}
F_{\mu\nu}^2 \ - \ Q_\mu D_\nu F_{\mu\nu} \ - \
\half F_{\mu\nu} [ Q_\mu , Q_\nu ]  
\] 
\be 
- \ \frac{1}{4}
G_{\mu\nu}^2 \ - \ \half G_{\mu\nu} 
[Q_\mu , Q_\nu ] \ - \ \frac{1}{4}
[ Q_\mu , Q_\nu ]^2 \ - \ \frac{i}{2} \pi ( \zeta_{YM} -
\eta_{YM})(A) \}
\la{correct1}
\ee
For  the instanton 
approximation (\ref{path3}) we introduce similarly
\[
Z_{YM} \ \approx \ \int [dA] [dQ] \delta 
( F^{\pm} ) \det ||
{ \delta F^{\pm} \over \delta A } || \exp \{ \int  - 
\frac{1}{4}
F_{\mu\nu}^2 \ - \ Q_\mu D_\nu F_{\mu\nu} \ - \
\half F_{\mu\nu} [ Q_\mu , Q_\nu ]
\]
\be
- \ \frac{1}{4}
G_{\mu\nu}^2 \ - \ \half G_{\mu\nu} 
[Q_\mu , Q_\nu ] \ - \ \frac{1}{4}
[ Q_\mu , Q_\nu ]^2 \ - \ \frac{i}{2} \pi 
( \zeta^{\pm}_{YM} -
\eta^{\pm}_{YM})(A) \}
\la{correct2}
\ee
where $\zeta_{YM}^{\pm}$ and $\eta_{YM}^{\pm}$ 
now denote the
$\zeta$-function and the 
$\eta$-invariant of the
(anti)selfdual fluctuation matrix. However, now
we do not necessarily expect to obtain
an improvement of (\ref{path3}). In fact, since
the fluctuation determinant in (\ref{path2}) is
non-negative for (anti)selfdual configurations, it
appears that when (\ref{correct2}) is summed over
both selfdual and anti-selfdual configurations
we reproduce the standard
instanton approximation of (\ref{path2}).

Since the fluctuation matrix of the original Yang-Mills
equation is an elliptic
second-order operator, the number of its negative 
eigenvalues is restricted and consequently (\ref{phases1})
is quite sufficient. However, since the fluctuation
matrices of the (anti)selfdual equations are first order
operators, the regulated expressions are more appropriate.

In general, the evaluation of the phase factors
$\zeta_{YM}$ and $\eta_{YM}$ 
is prohibitively complicated and consequently we
shall proceed with the more manageable
(\ref{path2}) and (\ref{path3}). But 
since the contribution
from these phase factors is additive, we conjecture that
the qualitative aspects of our analysis remain 
largely intact. Furthermore, in the following
we shall be mostly interested in the infrared limit
of the Yang-Mills theory.  In this limit we expect
both $\zeta_{YM}$ and $\eta_{YM}$ to become
irrelevant operators,
and we conjecture that these phase factors are
relevant only when we attempt
to compute higher order corrections.   

\vskip 0.3cm

\section{Topological Yang-Mills theory}

\vskip 0.2cm

We shall first investigate the 
(anti)selfdual approximation
(\ref{path3}). We wish to promote the
classical field $A_\mu$ into an 
off-shell field, by representing
the $\delta$-function and the determinant as a path
integral over a set of auxiliary variables. 
For this, we first need an appropriate
representation of the Euler character (\ref{GBC2A}).
For definiteness we specialize to the
anti-selfdual configurations $F^+ = 0$.

The path integral representation of the Euler
character
\be
\X( {\cal A}/{\cal G} ) ~=~
\sum\limits_{F^+ =0} sign \det || { \delta F^+ \over 
\delta A } ||
\la{GBC2}
\ee
has been investigated by 
Atiyah and Jeffrey \cite{atje}. They showed,
that (\ref{GBC2})
coincides with the partition function of
topological Yang-Mills 
theory \cite{witten} in the Mathai-Quillen
formalism \cite{bgv}. Their construction can be viewed as
a direct generalization of the familiar
result that the Euler character of a compact
Riemannian manifold can 
be represented by the partition
function of the de Rham 
supersymmetric quantum mechanics \cite{wit2},
\cite{blau}.

The partition function of topological Yang-Mills
theory is an example of a cohomological path integral
of the form
\be
Z_{TYM} ~=~ \int \exp \left( \{ \Omega_0 , 
\Psi \} \right)
\la{DW}
\ee
where $\Omega_0$ is a nilpotent BRST operator and $\Psi$
is a gauge fermion. Standard arguments imply that
this path integral describes only the cohomology
classes of $\Omega_0$, and it is formally
invariant under local variations of $\Psi$.

Atiyah and Jeffrey \cite{atje} constructed a 
one-parameter family of $\Psi$'s that interpolates 
between the Gauss-Bonnet-Chern
and the Poincar\'e-Hopf representatives of the
Euler class on ${\cal A}/{\cal G}$. In particular,
for a definite value of the parameter
their action reproduces
the original action of topological Yang-Mills theory
\cite{witten}. For this, we use the notation of Ouvry, 
Stora and van Baal \cite{osv} (except that we denote
by ${\bf u}$ and ${\bf v}$ the standard
ghosts for gauge fixing) and
introduce a graded symplectic manifold with
the following canonical variables

\vskip 0.5cm
\begin{center}
\begin{tabular}{|c|c|c|}
\hline
& & \\
form & ~~ EVEN ~~ & ~~ ODD ~~ \\ 
degree & (q,p) & (q,p) \\
& & \\
\hline 
& & \\
0-form ~~& $\varphi$ , $\pi$ & ${\bf u}$,  ${\bf v}$   \\
1-form ~~&  $A$, $E$      & $\psi$,   $\X$     \\
2-form ~~&  $b$, $c$      
& $\bar\psi$,  $\bar{\X}$ \\
4-form ~~& $\bar\varphi$ , 
$\bar\pi$ & $\beta$, $\gamma$ \\
& & \\ 
\hline
\end{tabular}
\end{center}
\vskip 0.5cm
The graded Poisson brackets of these variables are
\ba
\{ p^a , q^b \} ~ &=& ~ -\delta^{ab}
\\ \non
\{ p^a_\mu , q^b_\nu \} ~ &=& ~ - \delta^{ab}_{\mu\nu} 
\\ \non
\{ p^a_{\mu\nu} , q^b_{\rho\sigma} \} ~ 
&=& ~ - \frac{1}{4} \delta^{ab}
( \delta_{\mu\rho} \delta_{\nu\sigma} - 
\delta_{\mu\sigma}
\delta_{\nu\rho} + 
\epsilon_{\mu\nu\rho\sigma} )
\la{PB}
\ea 
where the 2-form bracket explicitly accounts 
for the antisymmetry and 
selfduality of the corresponding
variables.

The nilpotent BRST operator $\Omega_0$ that computes 
the cohomology of the topological Yang-Mills
theory can be represented as a linear combination 
\cite{osv}, \cite{maillet},
\be
\Omega_0 ~=~ \Omega_{TOP} \ + \ \Omega_{YM}
\la{brst1}
\ee
Here  
\be
\Omega_{YM} \ = \   {\bf u} \G \ + \ 
\half {\bf v} [{\bf u}, {\bf u}] \ + \ h \bar{\bf u}
\la{gauge1brst}
\ee
is the standard 
nilpotent BRST operator that fixes 
the SU(N) gauge invariance,
with 
\[
\G ~=~ D_\mu E_\mu + [\varphi , \pi ] + 
[ \bar\varphi , \bar\pi ]
+ [\beta , \gamma ] + [\psi_\mu , 
\X_\mu ] + [\bar\psi_{\mu\nu} , 
\bar\X_{\mu\nu} ] + [b_{\mu\nu} , 
c_{\mu\nu} ]
\]
the Gauss law operator which
generates the 
gauge transformations of the various fields,
\[
[ \G^a , \G^b ] ~=~ f^{abc} \G^c
\]
The other operator 
\be
\Omega_{TOP} \ = \ 
\psi_\mu E_\mu + \varphi ( D_\mu \X_\mu + 
[ \bar\psi_{\mu\nu} , c_{\mu\nu}] + {\bf v}) + 
b_{\mu\nu} \bar\X_{\mu\nu} + \gamma [ \varphi , 
\bar\varphi ]
+ \beta \bar\pi
\la{brst1top}
\ee
is the BRST operator for the topological 
symmetry. It is equivariantly nilpotent,
\[
\Omega_{TOP}^2 \ = \ - 2 \varphi {\cal G}
\]
Since this generates a gauge transformation
with $\varphi$ as the gauge parameter,
$\Omega_{TOP}$ is then nilpotent on the gauge 
orbit ${\cal A}/{\cal G}$.

From \cite{henne}, \cite{maillet} 
we conclude, that the {\it complete}
BRST operator of the topological Yang-Mills
theory is the linear combination
of (\ref{brst1}) with another nilpotent
operator $\Omega_{gf}$ which is necessary for
gauge fixing the various symmetries,
\be
\Omega_{full} ~=~ \Omega_0 \ + \ \Omega_{gf}
\la{ofull}
\ee
In (\ref{gauge1brst}) we have included
the pertinent $\Omega_{gf}$ that fixes the
standard SU(N) gauge symmetry, it corresponds
to the $h \bar{\bf u}$ term. 
In the general case the operators 
$\Omega_0$ and $\Omega_{gf}$ also
anticommute, ensuring the nilpotency
of $\Omega_{full}$. The structure of $\Omega_{gf}$
in the topological sector
has been explained {\it e.g.} in \cite{maillet},
and will be discussed in section 9 of the present paper.
Like the corresponding term 
in (\ref{gauge1brst}) it is trivial, 
and for the present purposes it is
sufficient to consider only $\Omega_0$.  
Consequently in the following 
we shall not consider $\Omega_{gf}$ explicitly in
the topological sector,
it is enough for us to assume
that the topological gauge invariances are fixed by
some appropriate gauge condition.

We construct the action of 
topological Yang-Mills
theory in the path integral (\ref{DW}), 
by first introducing four different
gauge fermions $\Psi_i$
\ba
\Psi_1 \ &=& \ \bar\psi \wedge b 
\non \\
\Psi_2 \ &=& \ \bar\psi \wedge F^+
\non \\
\Psi_3 \ &=& \ \star \bar\varphi \wedge D \star \psi
\non \\
\Psi_4 \ &=& \ \beta \wedge 
[ \varphi , \star \bar\varphi ]
\la{psis}
\ea
Here $\star$ denotes the Hodge duality 
transformation.
We introduce four numerical parameters
$\alpha_i$ and define the gauge invariant
topological action
\be
S_{TOP} \ = \ \sum\limits_{i=1}^{4} \
\alpha_i \{ \Omega_{TOP} , \Psi_i \}
\la{action} 
\ee
By substituting (\ref{psis}) and eliminating
the auxiliary field $b$ by
a Gaussian integration in (\ref{DW}), we obtain
\[
- S_{TOP} \ = \ \alpha_1  \varphi [ \bar\psi , 
\bar\psi ] \ + \ \alpha_2
\left[ - \frac{\alpha_2}{\alpha_1} 
\frac{1}{4} F^+ \wedge F^+ \ + \ 
D \psi^+ \wedge \bar\psi \right]
\]
\be
+ \alpha_3 \left( \star \beta 
\wedge D \star \psi + \star \bar\varphi
\wedge [ \psi , \star \psi ] - 
\star \bar\varphi D \star D \varphi \right)
+ \alpha_4 \left( [ \varphi , 
\star \bar\varphi ]^2 + \varphi [\star\beta ,
\star \beta ] \right) \star 1
\la{action2}
\ee
Different values of  $\alpha_i$ label
different representations of the 
theory, and by general arguments the ensuing
path integral (\ref{DW}) should be independent of
these parameters. For example, if we select 
\[
\alpha_1 \ = \ - \alpha_2 \ = \ \alpha_3 \ = \ 1
~,~~~~~~ \alpha_4 \ = \ 0
\]
we find the action presented 
in \cite{osv}. On the other
hand, the action originally 
introduced by Witten 
\cite{witten} emerges if we select
\be
\alpha_1 \ = \ 4 ~,~~~~~~ \alpha_2 
\ = \ - 4 ~,~~~~~~ \alpha_3 \ = \
1 ~,~~~~~~ \alpha_4 \ = \ \half
\la{parat1}
\ee
and use the
following identification of 
variables between the
notations in \cite{osv} and \cite{witten},

\vskip 0.5cm
\begin{center}
\begin{tabular}{|c|c|}
\hline 
& \\
~~ OSvB ~~ & ~~ Witten ~~ \\ 
& \\
\hline
& \\
A & A \\
$\psi$ & $i\psi$ \\
$\varphi$ & $i \phi$ \\
$\bar\psi$ & $\frac{1}{4} \X$ \\
$\star\bar\varphi$ & $- \frac{i}{2} \lambda$ \\
$\star\beta$ & $\eta$ \\
& \\
\hline
\end{tabular}
\end{center}
\vskip 0.5cm
The result is
\[
- S_{W} ~=~ - \frac{1}{4}  F_{\mu\nu}^2 
\ - \ \frac{1}{4} F_{\mu\nu} 
{\tilde F}_{\mu\nu} \ - \ \half
\phi D_{\mu}^2 \lambda \ + \  i \eta 
D_\mu \psi_\mu \ - \ i D_\mu \psi_\nu
\X_{\mu\nu}
\]
\be
+ \ \frac{i}{8} \phi [ \X_{\mu\nu} , 
\X_{\mu\nu} ]
+ \ \frac{i}{2} \lambda [ \psi_\mu , 
\psi_\mu ] + \ \frac{i}{2}
\phi [ \eta , \eta ] + \frac{1}{8} 
[\phi , \lambda ]^2
\la{SW}
\ee

In \cite{atje} Atiyah and Jeffrey showed, 
that the 
corresponding path integral
(\ref{DW}) yields the Euler character
(\ref{GBC2}) on ${\cal A}/{\cal G}$. 
For this, we return to 
the notation of \cite{osv}
and select
\[
\alpha_2 \ = \ -1 ~,~~~~~~
\alpha_4 \ = \ 0
\]
but leave $\alpha_1$ and $\alpha_3$ arbitrary.
We again eliminate $b$ by a 
Gaussian integration and
find for the path integral (\ref{DW})
\[
Z_{TYM}  =  \int [dA] \ ... \ [d\lambda] 
[ \sqrt{ {1 \over 4 \pi 
\alpha_{1}} } ]
\exp \{ - \int  \frac{1}{4\alpha_1} 
(F^+)^2  \ + \ D \psi^+ \wedge \bar\psi \ - \ 
\alpha_1 \varphi [ \bar\psi ,
\bar\psi ] 
\]
\be
- \  \alpha_3 ( \star\beta \wedge D \star 
\psi + \star \bar\varphi
\wedge [ \psi , \star \psi ] - \star 
\bar\varphi D \star D \varphi )
\}
\la{MQ1}
\ee
The parameter $\alpha_3$ can be 
eliminated by redefining
$\beta$ and $\bar\varphi$, and we get
\[
Z_{TYM}  =  \int [dA] [d\psi] 
[d\bar\psi] [d\varphi] [ \sqrt{ {1 
\over 4 \pi \alpha_1 } } ]
\exp \{ - \int  \frac{1}{4\alpha_1} 
(F^+)^2  \ + \ D \psi^+ \wedge 
\bar\psi 
\]
\be
- \  \alpha_1 \left( { 1 \over
\star D \star D } \star [ 
\psi , \star \psi ] \right) \cdot
[\bar\psi , \bar\psi ] \ \}
\la{MQ2}
\ee
Here
\be
R \ = \ { 1 \over \star D \star D } 
\star [ \psi , \star \psi ]
\la{MQR}
\ee
can be identified as the curvature 
two-form on ${\cal A}/{\cal G}$  when 
we restrict $\psi$ 
to be horizontal on the gauge orbit space, 
\be
D \star \psi \ = \ 0
\la{hori}
\ee
which follows as a $\delta$-function constraint
when we integrate over $\beta$ in (\ref{MQ1}).
(We assume that a proper gauge fixing - determined
by $\Omega_{gf}$ in (\ref{ofull}) - has been
introduced, so that the 
integral in (\ref{MQ2})
extends only over the gauge orbit 
${\cal A}/{\cal G}$.)
Indeed, if we introduce the connection
\be
\Gamma \ = \ { 1 \over \star D 
\star D } \star D \star
\psi
\la{christ}
\ee
and define the exterior derivative by
\[
d \ = \ \psi^a_\mu  { \delta \over \delta A^a_\mu }
\]
we find that the curvature two-form 
\[
R \ = \ d \Gamma \ + \ \Gamma \wedge \Gamma
\]
coincides
with (\ref{MQR}) when we restrict to the 
horizontal bundle (\ref{hori}).

According to general arguments, the path integral
(\ref{MQ2}) is at least formally independent
of $\alpha_1$ and we can
interpret it by considering various limits: 

When $\alpha_1 \to \infty$ 
we find  
that (\ref{MQ2}) reduces to 
the Gauss-Bonnet-Chern representation
of the Euler character on ${\cal A}/{\cal G}$,
\be
Z_{TYM} ~=~ \int [dA][d\psi]{\rm Pf} (R)
\la{tymec}
\ee
where $R$ is the curvature two-form (\ref{MQR}).
 
On the other hand, when 
$\alpha_1 \to 0$ we get
\be
Z_{TYM} \ = \ \int [dA] \delta (F^+) \det || D^+ ||
~ \approx ~ \sum\limits_{F^+ = 0} sign \det 
|| {\delta F^+ \over
\delta A } ||
\la{GBC}
\ee
which is the 
Poincar\'e-Hopf representation (\ref{GBC2})
of the Euler character on ${\cal A}/{\cal G}$. 
As a consequence we have a 
generalization of the finite dimensional
relation between the Gauss-Bonnet-Chern
and Poincar\'e-Hopf theorems, and in particular
(\ref{MQR}) is indeed the
curvature two-form on ${\cal A}/{\cal G}$.

\vskip 0.3cm

\section{The Instanton Approximation}

\vskip 0.2cm

We shall now apply the results of the 
previous section to investigate
the anti-selfdual 
approximation (\ref{path3}) to the Yang-Mills
partition function (\ref{SYM}),
\be
Z_{YM} \ = \ \sum\limits_{F^+ = 0 } 
sign \det || {\delta F^+ \over
\delta A } || \left\{ \int [dQ]
\exp\{ \int  - \frac{1}{4}  
F_{\mu\nu}^2 (A + Q) \} \ \right\}
\la{YMTYM1}
\ee
Evidently the selfdual approximation is similar,
and there is no need to consider it explicitly.

For this we define the following 
more general path integral
\be
Z_{YM}(\alpha_1, \alpha_3)  
\ = \ \int [dQ][dA] ... [d \beta ]
\exp \{ - S_{YM}(A + Q ) \ - \  
S_{AJ}(A;\alpha_1, \alpha_3) \}
\la{YMAJ}
\ee
where $S_{YM}$ is the $A_\mu + Q_\mu$ dependent
Yang-Mills background field 
action that appears in (\ref{YMTYM1}) 
and $S_{AJ}$ is the $A_\mu$ dependent
Atiyah-Jeffrey representation
of the topological
Yang-Mills action that appears
in (\ref{MQ1}).
Since the $\alpha_1 \to 0$ limit of
(\ref{MQ2}) localizes to the Poincar\'e-Hopf
representation (\ref{GBC}) of the Euler character, 
we conclude that
we obtain (\ref{YMTYM1}) when 
$\alpha_1 \to 0$ in (\ref{YMAJ}),
\[
\lim_{\alpha_1 \to 0} Z_{YM}(\alpha_1 ,
\alpha_3 ) ~ = ~ Z_{YM}
\]

A priori, the path integral 
(\ref{YMAJ}) depends nontrivially on the
parameters $\alpha_1$ and $\alpha_3$. However,
we shall now argue that (\ref{YMAJ}) is
actually {\it independent}
of $\alpha_1$ and $\alpha_3$ so that
it coincides with  (\ref{YMTYM1})
independently of these parameters.
More generally, we shall argue that independently
of the parameters $\alpha_i$ we can write
(\ref{YMTYM1}) as
\be
Z_{YM} ~=~ \int [dQ] [dA] ... [d\beta]
\exp \left(  \int - \frac{1}{4} 
F_{\mu\nu}^2(A+Q) \ - \ \sum_{i=1}^{4}
\ \alpha_i \{ \Omega_{TOP} , \Psi_i \} \ \right)
\la{YMTYM2}
\ee 
where $\Omega_{TOP}$ is the topological
BRST operator (\ref{brst1top})
and $\Psi_i$ are the gauge fermions defined in
(\ref{psis}). 

Notice in particular, that the second 
term in (\ref{YMTYM2}) 
depends only on the classical field $A_\mu$, and 
coincides with the action (\ref{action}) 
of topological Yang-Mills theory.

In order to demonstrate 
the $\alpha_i$ independence of
(\ref{YMTYM2}) we first introduce a new
variable $P^a_\mu$ which is conjugate 
to the fluctuation field $Q^a_\mu$,
\[
\{ P^a_\mu , Q^b_\nu \} ~=~ - 
\delta^{ab}_{\mu\nu}
\]
We then define the following linear combinations,
\ba
A^+_\mu \ &=& \ A_\mu + Q_\mu \non \\
A^-_\mu \ &=& \ A_\mu \non \\
E^+_\mu \ &=& \ P_\mu \non \\
E^-_\mu \ &=& \ E_\mu - P_\mu 
\la{newvar1}
\ea
Since the only nonvanishing Poisson brackets are
\be
\{ E^{\pm}_\mu, A^{\pm}_\nu \} \ = 
\ - \delta_{\mu\nu}
\la{newvarpb}
\ee
we conclude that $A^+_\mu$ and $A^-_\mu$ are actually
two {\it independent} gauge fields. 

More generally, 
we extend these $\pm$ variables
into a one-parameter family of $\pm$
variables by introducing
a canonical conjugation generated by
\[
\Phi ~ = ~ - \tau E_\mu Q_\mu
\]
where $\tau$ is a parameter. This gives 
\ba
A^{+}_\mu ~ \to ~ e^{ - \Phi } 
A^{+}_\mu e^{\Phi} ~ &=& ~ 
A_\mu + (1-\tau) Q_\mu \non \\
A^{-}_\mu ~ \to ~ e^{ - \Phi } 
A^-_\mu e^{\Phi} ~ &=& ~
A_\mu -\tau Q_\mu \non \\
E^+_\mu  ~ \to ~ e^{ - \Phi } 
E^+_\mu e^{\Phi} ~ &=& ~
P_\mu + \tau E_\mu \non \\
E^-_\mu ~ \to ~ e^{ - \Phi } 
E^-_\mu e^{\Phi} ~ &=& ~ 
- P_\mu + (1-\tau) E_\mu
\la{newvar2}
\ea
and reproduces (\ref{newvar1}) for $\tau = 0$.
Since this is a canonical transformation, 
the Poisson brackets
(\ref{newvarpb}) are preserved.
Notice in particular,
that independently of $\tau$
both fields $A_\mu^{\pm}$ gauge transform
like a SU(N) gauge field,
\[
A_\mu^{\pm} ~=~ U A_\mu^{\pm}U^{-1} + 
U \partial_\mu U^{-1}
\]

The present construction implies that the
action in (\ref{YMTYM2}) separates into two independent
contributions. The first term depends only on the
gauge field $A^+_\mu$ and the second term
depends only on the gauge field $A^-_\mu$,
\be
S_{YM}(A^+) + S_{TOP}(A^-)  ~=~  \frac{1}{4} 
F^2_{\mu\nu} ( A^{+}) \ + \ 
\sum\limits_{i=1}^{4} \alpha_i \{ 
\Omega^{-}_{TOP} , \Psi_i \} (A^-)
\la{sepact1}
\ee
where the BRST operator $\Omega^{-}_{TOP}$ now
acts only
on the fields $A^-$ and $E^-$,
\be
\Omega^{-}_{TOP} \ = \ \psi_\mu E^-_\mu + 
\varphi( D^-_\mu \X_\mu +
[\bar\psi_{\mu\nu} , c_{\mu\nu} ] + \rho ) + b_{\mu\nu}
{\bar \X}_{\mu\nu} + \gamma 
[\varphi , \bar \varphi ] + \beta \bar \pi
\la{brst2}
\ee
Here $D^-_\mu$ is covariant 
derivative {\it w.r.t.} the
gauge field $A_\mu^-$. 
For $\tau = 0$ (\ref{brst2})
reduces to (\ref{brst1}) 
except for the (irrelevant) shift
$E_\mu \to E_\mu - P_\mu$, and 
since $\tau$ only parametrizes 
a canonical transformation
this establishes the  asserted
$\alpha_i$-independence of the path integral
(\ref{YMTYM2}). In particular, 
by selecting $\alpha_2 = -1$,  $\alpha_4 = 0$ 
and taking the $\alpha_1 
\to 0$ limit we obtain our
anti-selfdual approximation (\ref{YMTYM1}).

Notice that the present construction also implies,
that instead of the original SU(N) 
gauge symmetry we now have two {\it independent} 
SU(N) gauge symmetries acting on the 
fields $A^\pm$ respectively, and
we have a SU(N)$\times$SU(N) gauge theory. 
However, the gauge fields $A^{\pm}_\mu$ are not entirely
independent. Since 
\be
A^+_\mu - A^-_\mu \ = \ Q_\mu
\la{trivbc}
\ee
obeys trivial boundary conditions the gauge
fields $A^{\pm}_\mu$ are subject
to the condition that their 
second Chern classes coincide,
\be
\int F_{\mu\nu} {\tilde F}_{\mu\nu} (A^+ ) ~ = ~
\int F_{\mu\nu} {\tilde F}_{\mu\nu} (A^- ) 
\la{bccc}
\ee
This also ensures that the path integral
in our anti-selfdual approximation,
\be
Z_{YM} ~=~ \int [dA^+][dA^- ] ... [d\beta] \exp
\{ \int - \frac{1}{4}F_{\mu\nu}^2 (A^+) \ - \ 
\sum\limits_{i=1}^{4} \alpha_i 
\{ \Omega^{-}_{TOP} , \Psi_i \} (A^-) \}
\la{YMTYM3}
\ee 
does not factorize into two independent
$\pm$ partition functions.
But since the Chern class only depends on the
global properties of the gauge field, the 
local {\it i.e.} perturbative fluctuations
of the fields $A^+$ and $A^-$ 
are independent. 
This implies
that in the perturbative 
sector parametrized by $Q$
we indeed have a {\it local} 
SU(N)${}_+\times$SU(N)${}_-$
gauge invariance. 

In order to eliminate
this SU(N)${}_+\times$SU(N)${}_-$
gauge invariance, we need a BRST operator
for both SU(N): In addition of 
the BRST operator (\ref{gauge1brst}) 
which eliminates the
SU(N)${}_-$ gauge invariance
\be
\Omega^{-}_{YM} \ = \   {\bf u} \G^{-} \ + \ 
\half {\bf v} [{\bf u}, {\bf u}] \ + \ h \bar{\bf u}
\la{gauge2brst}
\ee
where the Gauss law operator ${\cal G}^{-}$ is now
\be
{\cal G}^- \ = \ D^-_\mu 
E^-_\mu + [\varphi , \pi ] +
[\bar\varphi , \bar\pi ] 
+ [\beta , \gamma ] + [\psi_\mu ,
\X_\mu ] + [\bar\psi_{\mu\nu} , 
\bar\X_{\mu\nu} ] + [b_{\mu\nu} ,
c_{\mu\nu} ]
\la{gsslw}
\ee
we also introduce the following 
nilpotent BRST operator for the 
SU(N)${}_+$ gauge group
\be
\Omega^{+}_{YM} ~=~ {\bf c} D^+_\mu 
E^+_\mu \ + \ \half {\bf b} 
[{\bf c} , {\bf c} ] \ + \ t \bar{\bf c}
\la{brst3}
\ee
We then add the corresponding gauge fixing terms to
the action in (\ref{YMTYM3}),
\be
S_{YM}(A^+) \ + \ S_{TOP}(A^-) \ = \ 
\int  \frac{1}{4}F_{\mu\nu}^2 (A^+) \ + \ 
\sum\limits_{i=1}^{4} \alpha_i 
\{ \Omega^{-}_{TOP} , \Psi_i \}  
\ + \ \{ \Omega^{+}_{YM} , \Psi^{+} \}
\ + \ \{ \Omega^{-}_{YM} , \Psi^{-} \} 
\la{YMTYM4}
\ee
Here $\Omega^{-}_{TOP}$ and $\Omega^{-}_{YM}$ depend 
only on $A^-$, $E^-$ while 
$\Omega^{+}_{YM}$ depends only 
on $A^+$, $E^+$. Hence $\Omega^{+}_{YM}$ anticommutes
with both $\Omega_{TOP}$ and $\Omega^{-}_{YM}$
and the ensuing
path integral is invariant under local variations of
the gauge fermions $\Psi_i (A^-) , \ \Psi^{\pm} (A^{\pm})$. 

Notice that since the $\pm$-theories couple only
by nonperturbative boundary terms, in perturbation 
theory the quantum theory determined by
(\ref{YMTYM4}) coincides with the
standard Yang-Mills perturbation 
theory, and in particular the 
perturbative high energy 
$\beta$-functions of these
two theories coincide.
Only when nonperturbative 
instanton effects become relevant,
will (\ref{YMTYM4}) deviate from standard Yang-Mills
theory. Notice also that even 
though our boundary condition
(\ref{bccc}) breaks cluster decomposition 
nonperturbatively, perturbative 
Green's functions do obey
cluster decomposition.
This is consistent with the ultraviolet asymptotic
freedom which implies that in the high energy limit
the gauge fields are physical 
degrees of freedom. This
is also consistent with the (expected) infrared 
confinement of Yang-Mills theory, which implies
that in the infrared
limit we can not directly 
associate the gauge fields with
physical degrees of freedom.

Finally we observe that if 
we select $\tau = \half$ and $\alpha_1 = 4$
as in (\ref{parat1}), we find a particularly 
convenient representation of the Yang-Mills
part of the action (\ref{YMTYM4}): 
We first obtain
\be
- \frac{1}{4} F_{\mu\nu}^2(A^+) 
\ - \ \frac{1}{4} 
F_{\mu\nu}^2(A^-)
\ \to \
- \frac{1}{4} F_{\mu\nu}^2(A + 
\half Q) \ - \ \frac{1}{4} 
F_{\mu\nu}^2(A- \half Q )
\la{ymfs}
\ee
but since this is an even function in 
the fluctuation field $Q_\mu$, 
the linear and cubic terms in $Q_\mu$ 
disappear. In particular the 
term which is linear
in $Q_\mu$ and proportional 
to the Yang-Mills equation
of motion will be absent.
If we rescale $Q$ by a factor 
of $2$, we then have explicitly (excluding the
gauge fixing terms)
\[
- S_{YM} - S_{TOP} \ = \ - \half  F_{\mu\nu}^2 \ - \ 
\frac{1}{2}  G_{\mu\nu}^2 \ - \
F_{\mu\nu} [ Q_\mu , Q_\nu ] \ - \  \half
[ Q_\mu , Q_\nu ]^2
\ - \ \half 
\phi D_\mu^2 \lambda \ + \ i\eta D_\mu \psi_\mu 
\]
\be
- \ i \ D_\mu \psi_\nu
\X_{\mu\nu} \ + \ \frac{i}{8} \phi [ 
\X_{\mu\nu} , \X_{\mu\nu} ] \
+ \ \frac{i}{2} \lambda [ \psi_\mu , 
\psi_\mu ] \ + \ \frac{i}{2}
\phi [ \eta , \eta ] \ + \ \frac{1}{8} 
[\phi , \lambda ]^2 \ + \ 4 \pi^2 N
\la{ouract2}
\ee
Here the explicit covariant 
derivatives are now {\it w.r.t.} the
connection $A - Q$, and the 
notation coincides with that
in \cite{witten}. The $4 
\pi^2 N$ denotes the second Chern
class that appears in (\ref{SW}).

\vskip 0.3cm

\section{A gauge invariant mass scale}

\vskip 0.2cm

By construction, the action (\ref{YMTYM4}), (\ref{ouract2})
has a local
SU(N)${}_+\times$SU(N)${}_-$ gauge invariance.
However, the physically relevant gauge
invariance is the SU(N) symmetry of our 
original Yang-Mills theory. This suggests, that
physically interesting consequences such as the
possibility to introduce an explicit
gauge invariant mass scale
might emerge, if we allow the explicit breaking of the
SU(N)${}_+\times$SU(N)${}_-$ symmetry into the
original SU(N). From the
point of view of our original Yang-Mills theory
we may then view the ensuing SU(N) gauge theory, obtained
by explicitly breaking the SU(N)${}_+\times$SU(N)${}_-$
gauge symmetry into a SU(N),
as a natural 
extension of the original theory which may
have a number of physically desirable features.

Notice that in general an 
explicit gauge symmetry breaking
will also break the topological symmetry determined
by $\Omega^{-}_{TOP}$. As a consequence
the topological action will depend nontrivially 
on the parameters $\alpha_i$, which now become
coupling constants in our theory.

In order to explicitly break the extra SU(N) 
gauge symmetry, we observe that under the diagonal SU(N)
gauge transformations the linear combination
\be
{\cal A}_\mu ~=~ 
\frac{1}{2}( A^{+}_\mu + A^{-}_\mu ) ~=~ A_\mu +
\frac{1}{2} (1 - 2 \tau ) 
Q_\mu \ ~ {\buildrel 
{\tau = \frac{1}{2}} 
\over {\longrightarrow} } ~ \ A_\mu
\la{gaugef1}
\ee
transforms inhomogeneously 
like a gauge field, while the
linear combinations
\be
\Phi_\mu ~=~ A^{+} - A^{-} ~=~ Q_\mu
\la{higgs1}
\ee
transform homogeneously, 
like a Higgs field. This suggests, that we should
consider to
break the extra SU(N) symmetry by introducing
{\it e.g.} a gauge invariant mass term for the "Higgs
vector" $\Phi_\mu$. 
Consequently, instead of the original $A^{\pm}_\mu$ fields
it may be more convenient to represent
our action using the fields (\ref{gaugef1}), 
(\ref{higgs1}). 

In order to explicitly 
break the SU(N)$\times$SU(N) invariance,
we first consider the
BRST gauge fixing of these symmetries in our action.
For this we define the 
following two functionals,
\bastar
\Psi^+ \ & = & \ \bar{\bf v} ( \xi h +  
\partial_\mu A_\mu^- )
\non \\
\Psi^- \ & = & \ \bar{\bf b} ( \alpha t + 
D_\mu^- \Phi_\mu )
\eastar
and consider the ensuing BRST gauge fixing terms
\[
\{ \Omega_{YM}^+ , \Psi^+ \} \ + \ \{ 
\Omega_{YM}^- , \Psi^-
\}
\]
in the action (\ref{YMTYM4}). 
After we have eliminated the
auxiliary fields $t^a$, $h^a$ from the path 
integral by Gaussian integration 
we find the following
contribution,
\[
\frac{1}{4\xi} ( \partial_\mu A_\mu^-)^2 
\ + \ {\bf u} D_\mu^- \partial_\mu \bar{\bf v}
\ + \
\frac{1}{4\alpha} ( D_\mu^- \Phi_\mu )^2
\ + \ {\bf c} D_\mu^+ D_\mu^- \bar {\bf b} \ + \
\bar {\bf b} D_\mu^+ D_\mu^- {\bf u}
\]
Here we identify the first two terms
as the familiar
gauge fixing and  Faddeev-Popov
ghost terms in the
covariant $R_\xi$-gauge for $A_\mu^-$.
In addition, when $\alpha \to 0$ we 
find from the third
term the condition
\be
D_\mu^- \Phi_\mu \ = \ 0
\la{bggfc}
\ee
We can interpret this condition
as follows: 
The SU(N)${}_+$$\times$SU(N)${}_-$ gauge
invariance corresponds to the original
background formalism gauge invariance in (\ref{ym}).
The condition (\ref{bggfc}) fixes the gauge
invariance of the fluctuation field $\Phi_\mu
\sim Q_\mu$ with respect to the classical field
$A_\mu$ ($\sim A_\mu$ for $\tau = 0$), 
while the $R_\xi$ gauge condition
fixes the remaining gauge invariance of
the classical field $A_\mu$.

The condition (\ref{bggfc}) has also an alternative
interpretation. It can be
viewed as a unitarity condition for
the fluctuation field $Q_\mu \sim
\Phi_\mu$, since it eliminates the 
negative metric time component $\Phi_0$ 
in a covariant fashion. 

The present construction suggests that we
explicitly break the
SU(N)${}_+$$\times$SU(N)${}_-$ 
gauge invariance
into the SU(N) invariance of our original 
Yang-Mills theory
as follows:  
Since the field $\Phi_\mu$ transforms like 
a Higgs field (\ref{higgs1})
under gauge transformations and 
since (\ref{bggfc}) eliminates the unitarity
violating time component $\Phi_0$, we may 
view  $\Phi_\mu$ as a vector analog of 
the conventional scalar Higgs. Consequently we 
can explicitly break the 
SU(N)${}_+$$\times$SU(N)${}_-$ 
into the diagonal
SU(N) by adding the following "Higgs mass" to 
the action (\ref{ouract2}),
\be
S_m(\Phi) \ = \ m^2 \Phi_\mu^2
\la{higgsact}
\ee
Our action is then
\be
S \ = \ S_{YM} \ + \ S_{TOP} \ + \ S_m
\la{ouract3}
\ee

Since the SU(N)${}_+$$\times$SU(N)${}_-$ gauge
invariance has now been broken, we have also lost
the BRST invariances under (\ref{gauge2brst}) and
(\ref{brst3}). Hence we must construct
a new BRST operator for the remaining
SU(N) gauge symmetry. This BRST invariance must
also preserve the unitarity condition (\ref{bggfc}).
Since this unitarity condition is gauge covariant,
we conclude that the relevant nilpotent BRST operator
is
\be
\Omega_{diag} \ = \ {\cc}^a ( {\cal G}^a + 
f^{abc} {\uu}^b {\vv}^c )
+ \frac{1}{2} f^{abc} {\cc}^a {\cc}^b {\bb}^c + 
{\uu}^a D^{ab}_\mu \Phi^b_\mu + t^a {\bar {\cc}}^a + 
h^a {\bar {\uu}}^a
\la{brst2diag}
\ee
and our gauge fixed action is
\be
S \ = \ S_{YM} \ + \ S_{TOP} \ + \ S_m \ + \ \{ 
\Omega_{diag} , \Psi \}
\la{ouract5}
\ee
Indeed, consider the gauge fermion
\be
\Psi \ = \ \Psi_1 \ + \ \Psi_2  \ = \
\alpha \left( {\vv}^a 
D^{ab}_\mu \Phi^b_\mu \ + \  h^a \bar{\bf v}^a \right)
\ + \ {\bar {\bb}}^a ( 
\partial_\mu A^a_\mu - \frac{1}{4\xi} t^a )
\la{psi1s}
\ee
Here $\alpha$, $\xi$ are gauge parameters 
and the $\Psi$-invariance ensures that the
corresponding path integral is independent of these 
parameters. After we have eliminated $t^a$ by Gaussian
integration we find the following gauge fixed version
of (\ref{ouract5})
\[
- S \ = \ - \half  F_{\mu\nu}^2 \ - \ 
\frac{1}{2}  G_{\mu\nu}^2 \ - \
F_{\mu\nu} [ \Phi_\mu , \Phi_\nu ] \ - \
m^2 \Phi_\mu^2
\ - \  \half
[ \Phi_\mu , \Phi_\nu ]^2
\ - \ \half 
\phi D_\mu^2 \lambda 
\]
\[
+ \ i\eta D_\mu \psi_\mu \ - \ i \ D_\mu \psi_\nu
\X_{\mu\nu} \ + \ \frac{i}{8} \phi [ 
\X_{\mu\nu} , \X_{\mu\nu} ] \
+ \ \frac{i}{2} \lambda [ \psi_\mu , 
\psi_\mu ] \ + \ \frac{i}{2}
\phi [ \eta , \eta ] \ + \ \frac{1}{8} 
[\phi , \lambda ]^2 
\]
\be
- \ \alpha ( D_\mu \Phi_\mu )^2 \ - \
\xi (\partial_\mu A_\mu )^2 \ + 
\ \bar{\bf b}^a \partial_\mu
D^{ab}_\mu {\cc}^b \ + \ 4 \pi^2 N
\la{ouract6}
\ee
The $\alpha \to \infty$ limit then yields
the unitarity condition
\be
D^{ab}_\mu \Phi^b_\mu \ = \ 
\partial_\mu \Phi^a_\mu +
f^{acb} A^c_\mu \Phi^b_\mu \ = \
0
\la{subs1}
\ee
and the ${\bf b}$, $\bar{\bf c}$ ghost term 
coincides with the familiar Faddeev-Popov ghost 
term in the covariant
$R_\xi$ gauge for $A_\mu$.

The gauge fixed action (\ref{ouract6}) describes
the coupling of a massive vector field 
$\Phi_\mu$ to a SU(N) gauge field and the various
ghost fields of the topological Yang-Mills theory. 
Since the mass term for $\Phi_\mu$ breaks the 
topological BRST invariance under $\Omega_{TOP}$ 
in (\ref{YMTYM4}), these topological
ghost fields become physical. However, 
in the following section
we shall argue that the corresponding quantum theory is 
nonetheless unitary in the physical
subspace defined by demanding BRST invariance 
under (\ref{brst2diag}). Here we shall argue
that (\ref{ouract6}) is also power 
counting renormalizable.
For this, we observe that all 
interactions and ghost propagators
are power counting renormalizable. 
Consequently it is sufficient
to verify that the $A-\Phi$ 
propagators are also consistent
with power counting 
renormalizability {\it i.e.} that these
propagators vanish like ${\cal O}(p^{-2})$ in the 
$p^2 \to \infty$ limit.
Indeed, if for simplicity we rescale
\bastar
A_\mu & \rightarrow & \frac{1}{\sqrt{2} } A_\mu \\
\Phi_\mu & \rightarrow & \frac{1}{\sqrt{2} } \Phi_\mu
\eastar
we find for the propagators
\ba
\Delta^{AA}_{\mu\nu}(p) & = & {1 \over p^2} \left( 
\delta_{\mu\nu} - (1-{1
\over \xi}) {p_{\mu}p_{\nu} \over p^2} \right)  \non \\
\Delta^{\Phi\Phi}_{\mu\nu}(p) & = &  
\left( \delta_{\mu\nu} -
(1- \frac{1}{\alpha}) { p_{\mu}p_{\nu} \over p^2 + 
\frac{1}{\alpha} 
m^2} \right) {1 \over p^2 + m^2}
\ea
Here the $<AA>$-propagator has the standard 
form of a massless
gauge vector propagator in the covariant 
$R_\xi$-gauge. Similarly,
the $<\Phi\Phi>$-propagator has the 
standard form of a massive gauge vector
propagator in the covariant 
$R_\alpha$-gauge. Since both propagators
vanish like ${\cal O}(p^{-2})$ 
for large values of $p^2$ we
then conclude that our action is (at least)
power counting renormalizable, and the physical
spectrum contains both
a massless gauge field $A_\mu$ and a massive gauge field
$\Phi_\mu$.

\vskip 0.3cm

We conclude this section by noting, that (\ref{higgsact}) 
is not the most general SU(N) invariant
$\Phi_\mu$-dependent term
that one can introduce. However, (\ref{higgsact})
is particularly
interesting, since it introduces 
a power counting renormalizable mass scale
in a minimal and covariant fashion. More 
generally, one may consider
adding for example
\be
S_{\Phi} (\Phi) \ = \ \frac{1}{2} ( 
D_\mu \Phi_\nu )^2
\ + \ \frac{m^2}{2} \Phi_\mu^2 \ + \ \frac{g}{4} 
( \Phi_\mu^2 )^2
\la{higgsact2}
\ee
This coincides with the 
conventional form of a Higgs action
with four species of Higgs fields $\Phi_\mu$. This action
is also the most
general  one which is consistent with a twisted version 
of a Lorentz transformation where $\Phi_\mu$ transforms 
as a scalar instead of as a vector,
even though this twisted Lorentz invariance is broken by
the additional $ \Phi_\mu \sim Q_\mu $ 
dependent terms in (\ref{ouract2}). 
Indeed, the more general choice (\ref{higgsact2}) can 
be motivated by the {\it no-go} 
theorem derived in \cite{llew},
which states that if we are interested in introducing
a renormalizable mass for a gauge vector, 
a version of the  Higgs field is unavoidable. (However,
here we have also found that for power
counting renormalizability it is sufficient to consider
(\ref{higgsact}) only. This is due to the additional
$Q_\mu$ dependence which is present in our action.)

Using (\ref{higgsact2}), we can construct the following 
interesting and manifestly unitary alternative to our action
(\ref{ouract6}). For this  we first explicitly 
eliminate $\Phi_0 \sim Q_0$: 
Since $A^+$ and $A^-$ are independent
gauge fields, we can perform independent
SU$_{\pm}$(N) 
gauge rotations in (\ref{ouract2})
to gauge transform both fields to
the temporal gauge $A_0^{\pm} = 0$
before we explicitly break these 
gauge symmetries into the diagonal SU(N).
In terms
of our original fields, this means that we set
\[
A_0 \ = \ Q_0 \ = \ 0
\]
and in particular $\Phi_0 = 0$. In this gauge we 
can then explicitly break the remaining time-independent 
SU$_{+}$(N)$\times$SU$_{-}$(N) gauge invariance
into the diagonal SU(N), by introducing
the $\Phi_0=0$ version of 
(\ref{higgsact2})
\[
S_\Phi \ = \ \frac{1}{2}(D_\mu \Phi_i)^2 \ 
+ \ \frac{m^2}{2} \Phi_i^2 \ + \
\frac{g}{4}(\Phi_i^2)^2
\]
We now have a manifestly unitary theory,
albeit in a gauge which is not manifestly Lorentz-invariant.
However, this "gauge fixed" version of our
theory is particularly interesting, 
since we now
have three "scalar fields" $\Phi_i$ 
in the adjoint
representation of the gauge group. In analogy with standard
Yang-Mills-Higgs systems, we then expect that also in the
present case the
classical equations of motion admit finite energy 
magnetic monopole
solutions provided we select $m^2$ and $g$ so that 
we have a potential $V(\Phi)$ that exhibits
spontaneous symmetry breaking. Indeed, since these
solutions are time independent, by a gauge transformation
we can similarly eliminate one of the space
components of $\Phi_i$ and we are left with only two
"scalar fields" $\Phi_z$ and $\Phi_{\bar z}$.

Notice that since the "Higgs field" $\Phi_i$ by construction
obeys a trivial 
asymptotic boundary condition, we should only
consider monopole configurations that are consistent with
this boundary condition. This means that there must be
an equal number of monopoles and 
antimonopoles, and in particular we can
not have any "free" monopoles. In this sense
the present construction has certain aspects that are
reminiscent of
the picture of confinement introduced in \cite{thooft}.

\vskip 0.3cm

\section{Unitarity and N=2 supersymmetry}

\vskip 0.2cm

Since the action (\ref{ouract6}) 
fails to be invariant
under the BRST operator (\ref{brst2}), 
there is nothing  a priori that would prevent
the ghost fields of the topological Yang-Mills
theory from appearing in the physical 
spectrum. But since these ghosts violate
the spin-statistics theorem, this means that
unitarity is in peril. We 
shall now propose that the 
quantum theory of (\ref{ouract6}) 
is nevertheless unitary, by reformulating it
in terms of manifestly unitary variables.
For this we use the fact that the topological 
Yang-Mills theory can be viewed as a 
twisted version of the N=2 
supersymmetric Yang-Mills theory 
\cite{witten}, obtained by
re-interpreting the action of the Lorentz-group.
  
The (Minkowski space) action of 
the minimal SU(N) invariant
N=2 supersymmetric Yang-Mills theory is
\be
- S_{N=2} \ = \ -\frac{1}{4}  F_{\mu\nu}^2 - 
D_\mu B D_\mu \bar B
- i  \bar\lambda_{i} 
{\bar\sigma_{\mu}} D_\mu 
\lambda^{i} 
- \frac{1}{ \sqrt{2} }  B  
[ \bar \lambda_{i} , {\bar 
\lambda}^{i} ] +
\frac{1}{ \sqrt{2} }  
\bar B
[ \lambda_{ i } ,  
\lambda^{i} ] + \half [ B , 
\bar B ]^2 
\la{N=2}
\ee
By comparing the actions (\ref{SW}) and (\ref{N=2})
term-by-term, we observe an obvious similarity.
Indeed, since the Lorentz algebra SO(3,1) is related
to SO(4)$\sim$SU(2)$\times$SU(2)
and since the action (\ref{N=2}) 
has an internal SU(2)$^{N}$
symmetry, we can redefine the 
action of the Lorentz group
by twisting the fields. This means 
that we introduce an
invertible change of variables 
between (\ref{SW}) and (\ref{N=2})
\[
(B, \bar B, \lambda_i , \bar\lambda_i , A_\mu ) \ \to 
(\phi , \lambda , \eta , \psi_\mu , \X_{\mu\nu} , A_\mu )
\]
which  maps (\ref{N=2}) into (\ref{SW}) 
and {\it vice versa}.
This change of variables is defined by
\ba
B ~~ &=& ~~ - i \sqrt{2} \phi \non \\
\bar B ~~ &=& ~~- \frac{i}{\sqrt{8}} \lambda \non \\
\lambda_{\alpha i} ~~ &=& ~~ - \sigma_{\alpha i}^{\mu}
\psi^\mu \non \\
\bar\lambda_{\dot \alpha i } ~~ &=& ~~ - 
\half \epsilon_{\dot \alpha i} \eta + \frac{1}{4} 
{\bar\sigma^{\mu\nu}}_{\dot \alpha i} \X_{\mu\nu}
\la{variables1}
\ea
The inverse transformation is
\ba
\phi ~~ &=& ~~ \frac{i}{\sqrt{2}} B \non \\
\lambda ~~ &=& ~~ i \sqrt{8} \bar B \non \\
\eta ~~ &=& ~~ {\bar\lambda_i \hskip 0.05cm }^i \non \\
\psi_\mu ~~ &=& ~~ \half \sigma_\mu^{\alpha i} 
\lambda_{\alpha i} \non \\
\X_{\mu\nu} ~~ &=& ~~
2 {\bar\sigma_{\mu\nu}}^{\dot \alpha i} \bar\lambda_{\dot 
\alpha i}
\la{variables2}
\ea
If we substitute (\ref{variables1}) in the
N=2 action (\ref{N=2}) (modulo analytic continuation to
the Euclidean space and the 
topological $F \tilde F$ term) we find
\[
- S ~=~ - \frac{1}{4} F_{\mu\nu}^2 \ - \ 
\half \phi D_\mu^2 \lambda
\ + \ i\eta D_\mu \psi_\mu \ + \ i 
\X_{\mu\nu} D_\mu \psi_\nu
\ + \ \frac{i}{8} \phi [ \X_{\mu\nu} , 
\X_{\mu\nu} ]
\]
\be
\ + \ \frac{i}{2} \lambda [\psi_\mu , 
\psi_\mu ] \ + \ \frac{i}{2}
\phi [\eta , \eta] \ + \ \frac{1}{8} 
[\phi , \lambda ]^2
\la{notaa}
\ee
which is the action (\ref{SW}) of 
topological Yang-Mills theory.
Since this change of variables has 
a trivial Jacobian in the
path integral, we conclude that the ensuing
partition functions coincide. Only 
the interpretation
of these two theories is different.

Alternatively, we can introduce 
a change of variables which 
relates the N=2 theory to
the selfdual topological Yang-Mills theory  
instead of the anti-selfdual one.

Twisting also has an effect on the 
supersymmetry algebra:
For the N=2 theory we have the 
following N=2 supersymmetry
algebra,
\ba
\{ {Q_\alpha}^i , {\bar Q}_{\dot \beta j} \} 
\ &=& \ 2 {\delta^i}_j
\sigma^\mu_{\alpha \dot\beta} \ P_\mu 
\non \\
\{ {Q_\alpha}^i , {Q_\beta}^j \} \ &=& \ 
\epsilon_{\alpha\beta} 
\epsilon^{ij} Z 
\non \\
\{ {\bar Q}_{\dot \alpha i} , {\bar 
Q}_{\dot \beta j} \} \ 
&=& \ - \epsilon_{\dot \alpha 
\dot \beta} \epsilon_{ij} Z^{\star}
\la{susyalg1}
\ea
where $Z \sim a n_e + a_D n_m$ is the central charge.
Using our change of variables (\ref{variables1}), 
(\ref{variables2})
we then find, that the BRST 
operator (\ref{brst1}) of the
topological Yang-Mills theory is related to 
the N=2 supersymmetry generators of (\ref{N=2})
by
\be
\Omega \ = \ \epsilon^{\dot \alpha 
i} {\bar Q}_{\dot \alpha i}
\la{susyalg2}
\ee
Furthermore, since
\be
\{ {Q_{1}}^{2} , {\bar Q}_{\dot 1  2} \} \ 
= \ - 2 (H - P_3)
\la{1stsusyrel}
\ee
and
\be
\{ {Q_2}^{1} , {\bar Q}_{\dot 2 1} \} 
\ = \ - 2(H+P_3)
\la{2ndsusyrel}
\ee
where $H$ is the Hamiltonian of (\ref{N=2}), combining
(\ref{1stsusyrel}) and (\ref{2ndsusyrel}) we get
\[
H \ = \ - \frac{1}{4} \{ {\bar Q}_{\dot 1 2 } - 
{\bar Q}_{ \dot 2 1} , {Q_1}^2 - {Q_{2}}^1 \} \ = \ \{ 
\Omega , \Psi \}
\]
which implies that the gauge fermion $\Psi$ that yields
the action (\ref{notaa}) of topological Yang-Mills
theory is
\[
\Psi \ = \ \frac{1}{4} ( {Q_2}^1 - {Q_1}^2 )
\]

Finally, if we define
the following two operators
\bastar
Q^\mu \ &=& \ \sigma^{\mu}_{\alpha i} 
\epsilon^{\alpha\beta}
{Q_\beta}^i
\non \\
D^{\mu\nu} \ &=& \ { {\bar \sigma}^{\mu
\nu\dot\alpha} }_{\dot \beta}
\epsilon^{\dot \beta i} {\bar Q}_{\dot \alpha i}
\eastar
and introduce the antiself-dual projection operator
\[
{{\cal P}_-}^{\mu\nu\rho\sigma} \ = \
\frac{1}{4} \left( \delta^{\mu\rho} \delta^{\nu\sigma} - 
\delta^{\mu\sigma} \delta^{\nu\rho} - i
\epsilon^{\mu\nu\rho\sigma} \right)
\]
we find that in terms of the twisted variables
the supersymmetry algebra (\ref{susyalg1}) becomes
\bastar
\{ \Omega , \Omega \} \ &=& \ - 2 Z^{\star}
\non \\
\{ \Omega , Q^\mu \} \ &=& \ 4 P^\mu
\non \\
\{ \Omega , D^{\mu\nu} \} \ &=& \ 0
\non \\
\{ Q^\mu , Q^\nu \} \ &=& \ 2 \eta^{\mu\nu} Z
\non \\
\{ Q^\mu , D^{\rho\sigma} \} \ &=& \ 8 P_\nu {{\cal 
P}_-}^{\rho\sigma\mu\nu}
\non \\
\{ D^{\mu\nu} , D^{\rho\sigma} \} \ &=& \ - 2 Z^{\star}
{{\cal P}_-}^{\mu\nu\rho\sigma}
\eastar
Curiously, we find that if the central charge
$Z \sim a n_e + a_D n_m$ is
nonvanishing, the BRST operator 
becomes non-nilpotent. Since nilpotency is an essential
property of the BRST operator, 
this suggests an interesting
possibility to break the BRST 
symmetry in a dynamical fashion. It would be interesting
to understand, whether such a dynamical BRST symmetry
breaking could provide an alternative to our explicit
breaking of the BRST symmetries.

\vskip 0.2cm
We now return to the action (\ref{ouract6})
that we introduced in the previous section.
In terms of the N=2 variables (\ref{variables1}) we
find that this action becomes
\[
- S \ = \ - \ \half F_{\mu\nu}^2 \ - \ 
\frac{1}{2}  G_{\mu\nu}^2 
\ - \  F_{\mu\nu} [ \Phi_\mu , \Phi_\nu ] 
\ - \ m^2
\Phi_\mu^2 \ - \  \half
[ \Phi_\mu , \Phi_\nu ]^2 \ - \  D_\mu B D_\mu \bar B
\]
\be
- \ i  \bar\lambda_{i} {\bar\sigma_{\mu}} 
D_\mu \lambda^{i} \ + \ \half [ B , 
\bar B ]^2 \
- \ \frac{1}{ \sqrt{2} }  B  
[ {\bar \lambda}_{i } , {\bar \lambda}^{i} ] \ + \
\frac{1}{ \sqrt{2} }  \bar B 
[ \lambda_{i } ,  \lambda^{i} ] \ + \ 4\pi^2 N
\ + \ \{ \Omega_{diag} , \Psi \}
\la{actn2}
\ee
where the BRST operator is defined by (\ref{brst2diag}),
and it fixes the SU(N) gauge invariance of (\ref{actn2})
and also eliminates the nonunitary time
component of $\Phi_\mu$.
This action is consistent with the
spin-statistics theorem, and recalling our discussion
from the previous
section it also appears to be unitarity and power 
counting renormalizable. Thus we conclude that this massive
version of Yang-Mills theory appears to define a consistent
quantum field theory in four dimensions. In particular,
since 
(\ref{actn2}) is obtained from the standard Yang-Mills
theory simply 
by adding the mass term for the vector field $\Phi_\mu$,
it can be viewed as a natural massive generalization of
the standard Yang-Mills theory. Furthermore, since
the N=2 supersymmetric Yang-Mills theory which is contained
in (\ref{actn2}) has a very
rich structure and in particular confines \cite{nati},
the present extension may also imply physically  interesting
consequences for the ordinary Yang-Mills theory.

Notice that besides (\ref{higgsact}) 
there are also additional
gauge invariant, power-counting 
renormalizable and apparently
unitary mass terms that we can 
introduce. For example, following
\cite{nati} we could introduce a mass scale in (\ref{N=2}) 
by adding a mass term for the scalar superfield.
This mass term breaks the N=2 supersymmetry explicitly
into an N=1 supersymmetry, and as argued in \cite{nati}
in the context of the N=2 supersymmetry
the resulting theory exhibits both color and quark
confinement. In the present case, in
the absence of the explicit
SU(N)${}_+\times$SU(N)${}_-$ breaking 
terms such as (\ref{higgsact})
we then have an effective theory which couples this
confining N=1 theory to the original Yang-Mills theory only
by the non-perturbative fluctuations.

\vskip 0.2cm
We conclude this section by observing, that since
we are dealing with an effective theory obtained by summing
over anti-selfdual configurations in (\ref{YMTYM3}), there
is no a priori reason why our theory should
exhibit  perturbative
renormalizability. Unlike unitarity, 
perturbative renormalizability
is in general  
not expected to survive in an effective theory.
Rather, our arguments for renormalizability 
should be viewed
as a curiosity, that may point out new possibilities to
desribe massive gauge vectors without the Higgs effect. 
Furthermore, we recall
that (\ref{YMTYM3}) is only an approximation to the original
Yang-Mills path integral, and we should also 
account for the highly
nonlocal phase factors in (\ref{correct2}). These phase
factors break both perturbative renormalizability and
manifest locality.
However, since these phase factors represent an additive
structure in our theory, we do expect that 
the results derived in the present section,  
in particular the N=2 structure,  are
also present in the
original, exact version (\ref{SYM}) of the
Yang-Mills partition function.
Finally, since we are ultimately interested
in the infrared limit it is feasible to conjecture that
in this limit these highly complicated
phase factors become irrelevant
operators.

\vskip 0.3cm

\section{Quantum Morse theory and cohomology}

\vskip 0.2cm

In the previous construction
we have restricted our attention to
the anti-selfdual
solutions, approximating the
partition function (\ref{SYM}) by (\ref{path3}). 
We shall now consider the improvement (\ref{pathtwo})
that we obtain
when we sum over all possible
solutions to the Yang-Mills equations,
\be
Z_{YM} ~\approx~ \int [dQ][dA] \ \delta (DF)
\det || { \delta D F \over \delta A } ||
\exp \{ - \int \frac{1}{4}  F_{\mu\nu}^2(A+Q) \}
\la{allsol}
\ee
As we have explained in Section 2, we can formally
interpret this as an equivariant generalization
of the Euler character 
$\X({\cal A}/{\cal G})$ on the gauge orbit space,
\[
\X({\cal A}/{\cal G}) ~\approx~ \int [dA] \delta (DF)
\det || { \delta DF \over \delta A } ||
\]
\be
 =~ \sum\limits_{DF=0} sign \det || {\delta DF \over
\delta A } ||
\la{morseym}
\ee
In order to utilize this 
interpretation to study (\ref{allsol}), 
we first develop certain 
formal aspects. For this we consider
a generic D-dimensional quantum field theory defined by an
action $S(\phi^a)$, where $\{\phi^a \}$ are fields that take
values on some configuration space 
which in the case of Yang-Mills
theory is ${\cal A}/{\cal G}$.
For simplicity we shall assume that $S(\phi^a)$ has the
same formal properties as a nondegenerate Morse function.

Formally, according to the Poinc\'are-Hopf theorem
the Euler character of the
$\{\phi^a\}$ field space can be represented as
\[
\X (\phi ) \ = \ \sum\limits_{\delta S = 0} sign
\det || {\delta^2 S \over \delta 
\phi^a \delta \phi^b } || \ = \
\int [d\phi] \delta \left( {\delta 
S \over \delta \phi^a } \right)
\det || {\delta^2 S \over \delta 
\phi^a \delta \phi^b } ||
\]
\be
= \ \int [d\phi ] [d\pi ] [d \psi ] [ d{\bar{\cal P}} ]
\exp \{ i \int \pi^a {\delta S \over \delta \phi^a }
+ \psi^a {\delta^2 S \over \delta \phi^a \delta \phi^b }
\bar{\cal P}^b \} 
\la{morse1}
\ee
and a priori this is independent of the Morse function
$S(\phi)$. 
Here we have introduced one commuting ($\pi^a$) and two 
anticommuting ($\psi^a , \ 
\bar{\cal P}^a$) auxiliary variables,
to exponentiate the 
$\delta$-function and the determinant 
respectively. The action in (\ref{morse1})
\be
S_{eff} \ = \ \int \pi^a { \delta S \over \delta
\phi^a } \ + \ \psi^a { \delta^2 S \over \delta \phi^a 
\delta \phi^b } \bar{\cal P}^b 
\la{morseact}
\ee
has the following nilpotent supersymmetry
\ba
\Omega \phi^a      \ &=& \ \psi^a    \non \\
\Omega \psi^a      \ &=& \ 0         \non \\
\Omega \bar{\cal P}^a  \ &=& \ \pi^a \non \\ 
\Omega \pi^a   \ &=& \ 0
\la{pssusy1}
\ea
so that we can represent $\Omega$ by
\be
\Omega \ = \ \psi^a { \delta \over \delta \phi^a }
\ + \ \pi^a { \delta \over \delta \bar{\cal P}^a }
\la{delta}
\ee
and clearly
\[
\Omega^2 \ = \ 0
\]
We identify (\ref{pssusy1}) as the Parisi-Sourlas 
supersymmetry \cite{parisi}, when realized on a scalar 
superfield in the Parisi-Sourlas superspace. In 
addition of the space coordinates $x$, this superspace
has two anticommuting coordinates $\theta$ and $\bar\theta$ 
\[
\theta^2 \ = \ \bar\theta^2 \ = \ \theta \bar\theta 
+ \bar\theta \theta \ = \ 0
\]
The scalar superfield is
\be
\Phi^a(x,\theta,\bar\theta) \ = \ \phi^a + \theta \psi^a
- \bar\theta \bar{\cal P}^a + \theta \bar\theta \pi^a
\la{superfield}
\ee
and the supersymmetry (\ref{pssusy1}) can be identified with
the translation in the $\theta$ direction of the superspace,
\be
\Omega \ =  \  \int \partial_\theta 
\Phi^a {\delta \over \delta
\Phi^a } \ \sim \ \partial_\theta
\la{transl1}
\ee
Using (\ref{superfield}) we can write the action 
in (\ref{morse1}) as
\[
\int dx \ \left( \pi^a {\delta S 
\over \delta \phi^a }
\ + \ \psi^a {\delta^2 S \over 
\delta \phi^a \delta \phi^b }
\bar{\cal P}^b \right) \ = \ \int dx 
d \bar\theta d \theta \ 
S (\Phi)
\]
Hence, as a function of the 
superfield $\Phi^a$ 
the action (\ref{morseact})
has the same functional form as our 
original action $S(\phi)$.
Furthermore, if we introduce 
the following functional
\be
\Psi ~=~ \bar{\cal P}^a {\delta 
S \over \delta \phi^a }
\la{ekapsi}
\ee
we can represent (\ref{morseact}) 
as closed under the
supersymmetry operator (\ref{delta})
\be
\int \pi^a {\delta S \over \delta \phi^a }
+ \psi^a {\delta^2 S \over \delta 
\phi^a \delta \phi^b }
\bar{\cal P}^b \ = \ \int \Omega \Psi
\la{morseact1}
\ee
Consequently the path integral (\ref{morse1}) 
is of the standard
cohomological form (\ref{DW}),
\be
\X (\phi) \ = \ \int [d\phi][d\pi]
[d\bar{\cal P}][d\psi]
\exp \{ i \int \Omega \Psi \}
\la{integraali}
\ee
and in particular it is invariant under local variations
of the gauge fermion $\Psi$. Notice that this also ensures
that the Euler character (\ref{morse1}) is indeed
independent of the local details of
the ''Morse functional'' $S(\phi)$.

We shall now apply the $\Psi$-invariance 
of (\ref{integraali})
to generalize (\ref{ekapsi}) to
\be
\Psi \ = \ \bar{\cal P}^a ( {\delta S 
\over \delta \phi^a } + \frac{\kappa}{2}
\pi^a )
\la{uusipsi}
\ee
where $\kappa$ is a parameter. For the action this yields
\[
\int \Omega \Psi \ = \  \int \pi^a {\delta S 
\over \delta \phi^a }
\ + \ \psi^a {\delta^2 S \over \delta \phi^a \delta \phi^b }
\bar{\cal P}^b \ - \ \frac{\kappa}{2} \pi^2
\]
and in terms of the superfield $\Phi$ we get
\be
\int \Omega \Psi \ = \ \int dx d\bar\theta 
d \theta \left\{
S(\Phi) -  \frac{\kappa}{2} \Phi^a \partial_{\bar\theta} 
\partial_{\theta}
\Phi^a \right\}
\la{morseact2}
\ee
If the original action $S(\phi)$ has the standard
functional form
\be
S(\phi) \ = \ \half \phi^a ( - \Box ) \phi^a \ + \ V(\phi)
\la{convent}
\ee
we then conclude that the superspace action
can be represented in the corresponding
superspace form
\be
S(\Phi) \ = \ \half \Phi^a ( - \Box - 
\kappa \partial_{\bar\theta}
\partial_\theta ) \Phi^a \ + \ V(\Phi)
\la{psact1}
\ee
This is the standard Parisi-Sourlas 
action for a scalar field
theory that has been extensively investigated in
\cite{parisi}. In particular, it 
has been established that in perturbation
theory the superspace quantum field theory determined by
(\ref{psact1}) coincides diagram-by-diagram with
the quantum field theory of (\ref{convent}), but 
in two less space-time dimensions. 
This D $\to$ D-2 dimensional reduction
is a consequence of the negative dimensionality of
the anticommuting coordinates \cite{parisi}.
The dimensional transformation from the $D$ dimensional
coupling constants {\it etc.} to their 
D-2 dimensional counterparts is determined by the 
parameter $\kappa$ which has the dimensions 
\[
[\kappa] \ \propto \ m^2
\]
when we define the anticommuting variables $\theta$ and
$\bar\theta$ to be dimensionless. The overall numerical scale
of $\kappa$ is undetermined, and can be changed by redefining
the normalization of the $\theta$, $\bar\theta$ integral.

As explained in \cite{parisi}, the superspace quantum theory
can also be interpreted in terms of a stochastic differential
equation. For this we introduce an additional
variable $h$ and write the path integral (\ref{integraali}),
(\ref{uusipsi}) in the following
equivalent form
\[
\X(\phi) \ = \ \int [\frac{1}{2\sqrt{\kappa} }dh] 
[d\phi] \delta ( {\delta S \over
\delta \phi^a } - h^a ) \det || { 
\delta^2 S \over \delta \phi^a
\delta \phi^b } || \exp \{ - \int 
\frac{1}{4\kappa} h^2 \}
\]
\be
= \ \int [\frac{1}{2\sqrt{\kappa} }dh] 
\exp \{ - \int \frac{1}{4\kappa} h^2 \}
\sum\limits_{ {\delta S \over \delta \phi^a} = h^a }
sign \det || { \delta^2 S \over \delta 
\phi^a \delta \phi^b } ||
\la{random1}
\ee
This has the interpretation of 
averaging classical solutions
to the stochastic differential equation
\be
{\delta S \over \delta \phi^a} \ = \ - \Box \phi^a \ + \ 
\partial_a V(\phi) \ = \ h^a
\la{stocheqn}
\ee
over the external Gaussian random source $h$.
Notice that as a
consequence of the $\Psi$ invariance 
the integral (\ref{random1}) is actually independent
of $\kappa$, and if we take the $\kappa \to 0$ limit and
recall the Gaussian definition of a $\delta$-function
we find that (\ref{random1}) reduces to (\ref{morse1}). 
This is fully consistent with 
the fact that the Euler character 
is independent of the Morse functional. Indeed, from 
the Morse theory point of view
\[
S_h(\phi) \ = \ S(\phi) \ + \ h^a \phi^a
\]
is simply another (nondegenerate) Morse functional. (Here
we assume that the $h^a \phi^a$ 
term is a small perturbation
in the sense described in \cite{omat}.)

Finally, we conclude this section by deriving
the Gauss-Bonnet-Chern
theorem that represents $\X(\phi)$ in terms of the
curvature two-form on the configuration 
space $\{ \phi^a \}$.
(Here we assume that $\phi^a$ has a nontrivial
topology; If the configuration space is a
flat Euclidean manifold, see \cite{omat}). 
For this we introduce a canonical
transformation, determined by conjugating $\Omega$
\[
\Omega \ \to \ e^{-U} \Omega e^U 
\]
We select
\[
U \ =  \ - \Gamma^{a}_{bc} \psi^c \bar{\cal P}_a \lambda^b
\]
where we have introduced the conjugate variable
\[
\{ \pi^a , \lambda^b \} \ = \ - \delta^{ab}
\]
and $\Gamma^a_{bc}(\phi)$ are components of a connection
on the configuration space $\{ \phi^a \}$.
For the conjugated $\Omega$ we find the following 
transformation
laws,
\ba
\Omega \phi^a ~ &=& ~ \psi^a \non \\
\Omega \psi^a ~ &=& ~ 0 \non \\
\Omega \bar{\cal P}_a ~ &=& ~ \pi_a \ + \ \Gamma^c_{ab}
\psi^b \bar{\cal P}_c \\
\Omega \pi_a ~ &=& ~ \Gamma^c_{ab} \pi_c \psi^b \ - \ 
\half {R^c}_{adb} \psi^b \psi^d \bar{\cal P}_c \non 
\la{derham1}
\ea
which we identify as the familiar 
transformation law of the standard
(N=1) de Rham supersymmetric quantum mechanics \cite{palo}.
Indeed, if we assume that the 
connection $\Gamma^a_{bc}$ is metric
and use the metric tensor $g_{ab}(\phi)$ to define
\[
\Psi \ = \ g^{ab} \pi_a \bar{\cal P}_b
\]
we immediately 
find that the corresponding path integral (\ref{integraali})
evaluates to
\be
\X(\phi) \ = \ \int [d\phi][d\psi] 
\ {\rm  Pf} [ \half {R^a}_{bcd}
\psi^c \psi^d ]
\la{qcohom}
\ee
This is the formal (functional) 
Euler class of the configuration
space $\{ \phi^a \}$, and 
establishes that the Gauss-Bonnet-Chern 
representation of the Euler character 
indeed coincides
with the Poincar\'e-Hopf 
representation (\ref{morse1}),
(\ref{random1}).

\vskip 0.3cm

\section{All Solutions Approximation}

\vskip 0.2cm
We shall now study the approximation (\ref{pathtwo})
to the Yang-Mills partition function, using the formalism
we have developed in the previous section. 
We start by considering first
the formal Poincar\'e-Hopf 
representation of the Euler character
on ${\cal A}/{\cal G}$,
\[
\X ( {\cal A}/{\cal G} ) ~=~ \sum\limits_{DF=0} sign \det
|| { \delta DF \over \delta A } ||
\]
General arguments imply that this 
should coincide with (\ref{tymec}).
Indeed, by ignoring (irrelevant)
complications that arise from a nontrivial moduli
space we can show this directly, as a special
case of the construction we have
presented in the previous section. 
For this
we simply apply our derivation
of (\ref{qcohom}) to (\ref{morseym}), using the explicit
connection (\ref{christ})
and the corresponding curvature
two-form (\ref{MQR}) (and ignoring
inessential technical complications that arise since
(\ref{christ}) is not a metric connection). 
In this way we immediately find the
Gauss-Bonnet-Chern representation of the Euler 
character, consistent with (\ref{GBC2A}) and (\ref{tymec}).
This also generalizes the original
Atiyah-Jeffrey construction 
that we have presented in 
section 3, to the case where 
we account for all solutions to the
Yang-Mills equation. 

Here we are interested in the more general path 
integral (\ref{pathtwo}),
\[
Z_{YM} \ \approx \ \int [dQ][dA] \delta (DF) \det ||
{ \delta DF \over \delta A } || \exp 
\{  - \int \frac{1}{4}
F_{\mu\nu}^2 (A+Q ) \}
\]
which we interpret as an equivariant generalization
of the Euler character, as we have explained in section 2.
Following our discussion in
the previous section we introduce one commuting variable
$\pi_\mu$ and two anticommuting variables $\psi_\mu$ and
$\bar{\cal P}_\mu$ and write (\ref{pathtwo}) as
\[
Z_{YM} ~\approx~ \int [dQ][dA][d\pi][d\psi][d\bar{\cal P}]
\exp \{ \int - \frac{1}{4}  F_{\mu\nu}^2(A+Q) \ + \
\pi_\nu D_\mu F_{\mu\nu} \ - \
\bar{\cal P}_\mu [ F_{\mu\nu} , \psi_\nu ] 
\]
\be
- \ \half ( D_\mu \psi_\nu - D_\nu \psi_\mu )
( D_\mu \bar{\cal P}_\nu - D_\nu \bar{\cal P}_\mu ) \ \} 
\la{all1}
\ee
Here the first term depends 
on $A+Q$, while the remaining terms
depend on $A$ only. Consequently we can again introduce
the independent $\pm$ gauge fields 
(\ref{newvar2}) to conclude that the action
in (\ref{all1}) separates into a linear
combination of an $A^+ \sim A+Q$ dependent action
described by the first term in (\ref{all1}),
and an $A^- \sim A$ dependent
action described by the remaining terms in
(\ref{all1}). As we have explained in
section 5, this also means that 
we have a local
SU(N)${}_+\times$SU(N)${}_-$  gauge symmetry, and the
$\pm$ terms are coupled to each other only by
the requirement that the second Chern classes
coincide,
\[
\int F \tilde F (A^+) \ = \ \int F \tilde F (A^-)
\]

Consider the $A^-$ dependent, 
topological part of the action in (\ref{all1}),
\be
S_{TOP}(A^-) ~=~  \pi_\nu D_\mu F_{\mu\nu} \ - \
\bar{\cal P}_\mu [ F_{\mu\nu} , \psi_\nu ] \ 
- \ \half ( D_\mu \psi_\nu - D_\nu \psi_\mu )
( D_\mu \bar{\cal P}_\nu - D_\nu \bar{\cal P}_\mu )
\la{all2}
\ee
As we have explained in the
previous section, this action admits the following 
nilpotent BRST (Parisi-Sourlas) symmetry
\ba
\Omega A^-_\mu ~ & = & ~ \psi_\mu \non \\
\Omega \psi_\mu ~ &=& ~ 0
\non \\
\Omega \bar{\cal P}_\mu ~ & = & ~ \pi_\mu \non \\
\Omega \pi_\mu ~ &=& ~ 0
\la{all3}
\ea
so that
\be
\Omega \ = \ \psi_\mu E^-_\mu \ + \ 
\pi_\mu {\bar \eta}_\mu
\la{psbrst}
\ee
where we have introduced the conjugate variable
\[
\{ {\bar {\cal P}}_\mu , \bar\eta_\nu \} \ 
= \ - \delta_{\mu\nu}
\]
In particular, the action (\ref{all2})
can be represented as a BRST commutator,
\be
- S_{TOP} ~=~ \{ \Omega ,  \bar{\cal P}_\mu D_\nu F_{\mu\nu} \}
\la{comtop}
\ee
Furthermore, if we introduce the space components
of the Parisi-Sourlas supergauge field
\be
{\cal A}_\mu \ = \ A^-_\mu + \theta \psi_\mu \ -
\bar\theta \bar{\cal P}_\mu + \theta \bar\theta \pi_\mu
\la{spacesuperA}
\ee
we conclude from the previous section that
we can write (\ref{all2}) as
\be
S_{TOP} ({\cal A}) \ = \ \frac{1}{4}
\int dx d\bar\theta d\theta
\ {\cal F}_{\mu\nu}^2
\la{psym1}
\ee
where ${\cal F}_{\mu\nu}$ denotes the space-time
components of the Parisi-Sourlas field
strength tensor,
\be
{\cal F}_{\mu\nu} \ = \ \partial_\mu {\cal A}_\nu -
\partial_\nu {\cal A}_\mu + [ {\cal A}_\mu , {\cal A}_\nu ]
\la{psf1}
\ee

Since $\Omega$ does not act on the $A^+$ field, the
BRST symmetry (\ref{all3}) is also an invariance of the
full action in (\ref{all1}). Consequently
the path integral
\be
Z_{YM} ~\approx~ \int [dQ][dA][d\pi][d\psi][d\bar{\cal P}]
\exp \left\{ \int - \frac{1}{4}  F_{\mu\nu}^2(A^+)\ - \ 
\{ \Omega , \Psi \} (A^-) \ 
\right\}
\la{pathps}
\ee
is invariant under local variations of the gauge fermion
$\Psi$, and reproduces (\ref{all1}) when we select
$\Psi$ as in (\ref{comtop}).

In analogy with the previous section,
we now consider the following more general
gauge fermion
\be
\Psi ~=~  \bar{\cal P}_\mu ( D_\nu F_{\mu\nu} 
+ \kappa \pi_\mu )
\la{moregen}
\ee
where $\kappa \propto m^2$ is a mass scale.
Standard arguments imply
that the ensuing path integral (\ref{pathps}) is formally
independent of $\kappa$, and when  $\kappa \to  0$ we recover
(\ref{comtop}).

With nontrivial $\kappa$, 
the topological part of the action becomes
\be
\{ \Omega , \Psi \} ~=~  \pi_\nu D_\mu F_{\mu\nu} \ + \
\bar{\cal P}_\mu [ F_{\mu\nu} , \psi_\nu ] \ 
+ \  \half ( D_\mu \psi_\nu - D_\nu \psi_\mu )
( D_\mu \bar{\cal P}_\nu - D_\nu \bar{\cal P}_\mu ) \ 
+ \  \kappa \pi_\mu^2
\la{psact2}
\ee
To interpret this, we introduce the full Parisi-Sourlas
Yang-Mills field strength
\[
{\cal F}_{\alpha \beta} ~=~ 
\partial_\alpha {\cal A}_\beta \ - \ 
\partial_\beta {\cal A}_\alpha \ + \ [ {\cal A}_\alpha ,
{\cal A}_\beta ]
\]
where $\alpha, \beta = \mu, \theta, \bar\theta$. We then
define the full Yang-Mills action in the Parisi-Sourlas
superspace,
\be
S_{PS} ~=~ \int d\theta d\bar\theta \ \frac{1}{4}
{\cal F}^{\alpha \beta} {\cal F}_{\beta \alpha}
\la{psact3}
\ee
If we evaluate this action in the special case where
\[
{\cal A}_{\theta} \ = \ {\cal A}_{\bar\theta} \ = \ 0
\]
we find (\ref{psact2}) by integrating
over $\theta$ and $\bar \theta$. Hence we conclude that
the topological action (\ref{all2}) in our path integral
(\ref{all1}) determines a 4+2 dimensional 
Parisi-Sourlas Yang-Mills theory. Indeed, in the next
section we shall argue, that ${\cal A}_{\theta}$ and
${\cal A}_{\bar\theta}$ can be identified
with the ghosts that we need for a complete gauge
fixing in the topological sector. 

In analogy with
(\ref{ouract5}) we can now break the 
SU(N)${}_+\times$SU(N)${}_-$ gauge symmetry explicitly
into the physical SU(N) {\it e.g.} by adding an explicit
mass term to the fluctuation field $Q_\mu$. Indeed, a
breaking of our extra symmetries 
becomes necessary if we want
our construction to have nontrivial physical
consequences:  This gauge symmetry breaking also
breaks the BRST symmetry in the
topological sector and ensures that the $\kappa$ dependence
in (\ref{psact2}) becomes nontrivial. 

For this, following
section 5 we introduce
\be
S_{full} \ = \ \int \frac{1}{4} F^2_{\mu\nu} (A^+) \ + \
\frac{m^2}{2} Q_\mu^2 \ + \ 
\int \frac{1}{4} {\cal F}_{\alpha\beta}^2
({\cal A})
\la{finalfull1}
\ee
As we have explained in section 5 unitarity
is ensured by the
horizontality condition
\[
D_\mu Q_\mu \ = \ 0
\]
and since the mass term
breaks the topological 
(Parisi-Sourlas) BRST invariance,
the action (\ref{finalfull1}) 
depends nontrivially also on the parameter
$\kappa$ which is necessary for physically
nontrivial consequences. 

It is interesting to consider our
theory (\ref{finalfull1}) 
in its topological sector: If we select the 
superspace analog of Feynman gauge,
we find that
the propagator of the superfield (\ref{spacesuperA})
is
\be
<{\cal A}_\mu {\cal A}_\nu> \ = 
\ \delta_{\mu\nu} \left( \ 
{ 1 \over p^2 } \ + \ { \kappa \bar\alpha
\alpha \over p^4 } \ \right)
\la{confprop}
\ee
which exhibits the infrared ${\cal O}(p^{-4})$ behavior 
that leads to a linear potential between two static sources.
This ${\cal O}(p^{-4})$ 
infrared behavior in (\ref{confprop}) is 
unique in the following sense.
By demanding locality and gauge 
invariance, we can generalize
the gauge fermion in (\ref{moregen}) by expanding
it in derivatives of $\pi_\mu$ as follows,
\[
\Psi \ = \ \bar{\cal P}_\mu( D_\nu F_{\mu\nu} + \kappa \pi_\mu
+ \kappa_1 D_\nu \pi_\nu \pi_\mu + \kappa_2 
D^2 \pi_\mu + ... )
\]
In the infrared limit $p \to 0$ we then 
conclude that the
dominant contribution to the propagator indeed
comes from (\ref{moregen}).

The infrared behaviour of (\ref{confprop}) 
suggests that in the topological sector our 
theory confines. Indeed, previously it has been
conjectured \cite{kopis} that in some sense the
large distance limit of Yang-Mills
vacuum can be viewed as a medium of randomly
distributed color-electric and color-magnetic fields. 
This means that in the infrared limit Yang-Mills
theory can be approximated by the following set
of equations
\ba
D^{ab}_\mu F^{b}_{\mu\nu} \ &=& \ h^a_\nu
\non \\
<h^a_\mu (x) h^b_\nu (y) > \ &=& \ \delta_{\mu\nu}^{ab}
\la{niol}
\ea
where the white noise random source $h_\mu^a$ describes
the random color-electric and color-magnetic vacuum
medium; Notice that since gauge invariance implies
\be
D_\mu D_\nu F_{\mu\nu} \ = \ 0
\la{lush}
\ee
for consistency we must interpret the equations (\ref{niol})
to be defined on the gauge orbit 
${\cal A}/{\cal G}$.
The argument
presented in \cite{kopis} states, that as a
consequence of
the Parisi-Sourlas mechanism the
equations (\ref{niol}) imply an effective dimensional
reduction D=4 $\to$ D=2 in the infrared limit
with the ensuing confinement of color.
Indeed, by a direct computation one can show 
that in the Parisi-Sourlas theory planar
Wilson loops 
obey an area law, with string tension $\sigma$ determined 
by $\kappa$
\[
\sigma \ = \ {1 \over 4 \pi} \kappa N g^2
\]
This is a
direct consequence of the Parisi-Sourlas 
dimensional reduction, that relates (\ref{psact2}) 
to the corresponding two-dimensional, ordinary 
Yang-Mills theory. 

We observe, that the equations (\ref{niol}) are exactly those
that we have derived earlier in (\ref{stocheqn}), 
when we take into account the inessential complication
(\ref{lush}) and interpret these equations on
the gauge orbit ${\cal A}/{\cal G}$
and with a nontrivial moduli. Consequently we
can view the present construction as a first principle
derivation of
the equations (\ref{niol}), suggesting
that the qualitative picture developed in \cite{kopis}
might indeed be a proper way to describe Yang-Mills theories
in the infrared limit. However, to ensure 
that the dependence on
the parameter $\kappa$ is nontrivial, in addition
we need to break the BRST symmetry of the topological sector 
{\it e.g.} by adding a mass term as in (\ref{finalfull1}).

\vskip 0.3cm

\section{Comparison of the Approaches}

\vskip 0.2cm

In the previous sections we have
studied the Yang-Mills partition function
in two different approaches. We have first investigated
the instanton
approximation (\ref{YMTYM1}), which yields an effective
description based on the topological
and the N=2 supersymmetric 
Yang-Mills theories. We have 
then considered the more general
case (\ref{all1}), where we sum over
all possible solutions to the classical Yang-Mills
equation. The general arguments that we have presented in
Section 2. suggest that these two approaches
should essentially coincide. For this
reason it is
of interest to compare these approaches.
In this way we can also expect to
learn how good the instanton approximation actually is.

In addition of the conjugate Yang-Mills
variables $A_\mu$ and $E_\mu$ we introduce the
following pairs of canonically conjugated 
variables
\be
\{ \X_\mu , \psi_\nu \} \ = \ 
\{ \bar\X_\mu , \bar\psi_\nu
\} \ = \ \{ \pi_\mu , \lambda_\nu \} \ = \ - 
\delta_{\mu\nu}
\la{brackets1}
\ee
where we now use the notation in \cite{maillet}. We 
combine our variables into 
the Parisi-Sourlas
superfields as follows - see also (\ref{spacesuperA})
\ba
{\cal A}_\mu (y) \ &=& \ A_\mu + \theta \psi_\mu + 
\bar\theta \bar\X_\mu - \theta \bar\theta \pi_\mu
\non \\
{\cal E}_\mu (y) \ &=& \  \lambda_\mu + 
\theta \bar\psi_\mu
+ \X_\mu \bar\theta + \theta \bar\theta E_\mu
\la{psvar1}
\ea
so that we have the superbrackets
\[
\{ {\cal E}_\mu (y_1) , {\cal A}_\nu (y_2) \} \ = \ - 
\delta_{\mu\nu} (y_1 - y_2)
\]
in the Parisi-Sourlas superspace. 
In order to construct the remaining $\theta$, $\bar\theta$
components, we consider the nilpotent operator 
(\ref{psbrst}), 
\be
\Omega \ = \ \psi_\mu E_\mu \ + \ \pi_\mu {\bar \psi}_\mu
\la{psbrst2}
\ee
We identify this operator as the BRST operator for the
constraint
\be
E_\mu \ \approx \ 0
\la{noelec}
\ee
However, these constraints (\ref{noelec}) are not independent
but are subject to Gauss law,
\be
D_\mu E_\mu \ \approx \ 0
\la{nytgauss}
\ee
which projects (\ref{psbrst2}) to the gauge orbit space 
${\cal A}/{\cal G}$.  Such a linear
relation among the constraints (\ref{noelec}) means
that the constraint algebra is reducible \cite{henne}.
 
To implement Gauss law as a reducibility
condition in the BRST operator (\ref{psbrst2}), 
we introduce the following
canonically conjugated Parisi-Sourlas superfields,
\ba
{\cal A}_{\theta} \ &=& \ \eta - \theta \varphi - 
\bar\theta \lambda + \theta \bar\theta \bar\eta
\non \\
{\cal E}_{\bar\theta} \ &=& \ {\bar {\cal P}} + \theta
\pi - \bar\theta p + \theta \bar\theta {\cal P}
\la{psvar2}
\ea
and
\ba
{\cal A}_{\bar\theta} \ &=& \ \bar b + \theta\bar\pi
+ \bar\theta b + \bar\theta \theta \rho 
\non \\
{\cal E}_{\theta} \ &=& \ l + \theta c + \bar\theta 
\bar\lambda + \bar\theta \theta \bar c
\la{psvar3}
\ea
where we use the notation in \cite{maillet}.
We introduce $g_{\mu\nu} = \delta_{\mu\nu}$ and
$g_{\theta \bar\theta} = - g_{\bar\theta \theta} = 1$
and define the brackets of the various component
fields in (\ref{psvar2}), (\ref{psvar3})
so that our superspace variables obey
the following Poisson brackets
\[
\{ {\cal E}_{\alpha} (y_1) , {\cal A}_{\beta} (y_2) \}
\ = \ - g_{\alpha \beta} \delta(y_1 - y_2)
\]
As explained in \cite{maillet},
the components of these additional superfields
can be identified as the various
ghosts that we need to introduce for a fully gauge
fixed quantization of the constrained system
(\ref{noelec}), (\ref{nytgauss}), according to the
Batalin-Fradkin algorithm \cite{henne}.
Consequently we generalize the BRST operator
(\ref{psbrst2}) into
\[
\Omega \ \to \ \psi_\mu E_\mu + \bar\psi_\mu \pi_\mu
+ \varphi {\cal P} - \bar\eta \pi - c \rho + \bar c
\bar \eta
\]
\be
= \ \int d\bar\theta d\theta \ g^{\alpha \beta}
\partial_{\theta} {\cal A}_{\alpha} {\cal E}_{\beta} \ \sim
\ \partial_{\theta}
\la{psbrst3}
\ee
which reproduces (\ref{transl1}) on our variables.

If we now introduce the following conjugation of (\ref{psbrst3})
\[
\Omega \ \to \ e^{-\Phi} \Omega e^{\Phi} \ = \ 
\Omega + \{ \Omega , \Phi \} + \frac{1}{2!} \{ \{ \Omega ,
\Phi \} , \Phi \} \ + \ ...
\]
where we select
\[
\Phi \ = \ \int d\bar\theta d\theta \ \theta (
{\cal A}^a_\theta {\cal D}^{ab}_i {\cal E}^b_i + \frac{1}{2} 
f^{abc} {\cal A}^a_\theta {\cal A}^b_\theta  {\cal 
E}^c_{\bar\theta} )
\]
where the covariant derivative ${\cal D}^{ab}_i$
is with respect
to the Parisi-Sourlas superconnection,
we find for the conjugated $\Omega$
\be
\Omega \ = \ \int ( g^{\alpha \beta} \partial_\theta
{\cal A}^a_\alpha {\cal E}^a_\beta + {\cal A}^a_\theta
{\cal D}^{ab}_i {\cal E}^b_\mu - \frac{1}{2} f^{abc}
{\cal A}^a_\theta {\cal A}^b_\theta {\cal E}^c_{\bar\theta} )
\la{psbrst4}
\ee
Here the first term coincides with the translation operator
(\ref{psbrst3}) in the $\theta$-direction, and the two
remaining terms have the standard form of a nilpotent
BRST operator 
for the superspace gauge transformation, with
\[
{\cal G} \ = \ {\cal D}_\mu {\cal E}_\mu
\]
the superspace
Gauss law operator and ${\cal A}_\theta$, ${\cal 
E}_{\bar\theta}$ viewed as the superspace ghost.

\vskip 0.3cm
We are now in a position to compare the anti-selfdual
instanton approximation (\ref{YMTYM1}) to (\ref{all1}) 
which accounts for all possible solutions 
to the classical Yang-Mills equation. 
For this, we substitute our
component field expansions (\ref{psvar1}), (\ref{psvar2})
and (\ref{psvar3}) in the BRST operator 
(\ref{psbrst4}). Comparing with the
BRST operator (\ref{brst1}) of topological
Yang-Mills theory we then find
that the functional forms of 
(\ref{psbrst4}) and (\ref{brst1})
coincide,  term-by-term.
The only difference between these two operators comes
from the anti-selfduality of the two-form 
variables in (\ref{brst1}),
which have been replaced by vectors 
in (\ref{psbrst4}): The anti-selfdual tensor
has three independent components, while a vector
has four. Consequently this replacement means
an extension of the pertinent complex.
Such an extension is natural, since the
full Yang-Mills equation contains
both the anti-selfdual and
selfdual equations. 
The extenxion then simply reflects the
fact that while (\ref{psbrst4}) accounts for 
all possible solutions to the Yang-Mills equation,
in (\ref{brst1}) we only consider the anti-selfdual
ones. Indeed, a restriction of (\ref{psbrst4})
to the anti-selfdual sector yields immediately 
the BRST operator
(\ref{brst1}).

From the preceding discussion we conclude that
our two approaches   
(\ref{path2}) and (\ref{path3}) {\it essentially coincide}: 
Except for the selfduality condition, the only additional
difference 
arises from the choice of the gauge fermion $\Psi$. However, 
due to the $\Psi$-independence of the 
partition function this difference is entirely irrelevant. 

Since the
topological Yang-Mills theory is related to the N=2 
supersymmetric Yang-Mills theory by the
invertible change of variables (\ref{variables1}),
the Parisi-Sourlas Yang-Mills
theory must also contain the N=2 
theory, {\it both} in
its selfdual and antiselfdual subsectors. 
This means in particular, that
the confinement mechanism which has been identified
in the N=2 theory \cite{nati} is also contained in
the topological Parisi-Sourlas
sector of ordinary Yang-Mills theory. 
Thus there
should be a direct relation between the picture of
confinement by monopole condensation
developed in \cite{thooft},
\cite{nati} and the picture of confinement by randomly
fluctuating color-electric and color-magnetic
fields developed in \cite{kopis}.  
Furthermore, the derivation \cite{kupi} that
planar Wilson loops in the Parisi-Sourlas theory 
obey an area law should also be directly applicable to the
corresponding Wilson loops in the N=2 approach, modulo
a change of variables that originates from the different
choice of gauge fermions $\Psi$.

Finally, in our discussion of (\ref{YMTYM1}) and
(\ref{all1}) we have ignored the phase factors
that appear in (\ref{correct1}) and (\ref{correct2}).
However, we do not expect these phase factors to modify
the qualitative aspects of our results: Since these
phase factors depend only on the topological connection
$A^{-}$, they do not break the
SU$_{+}$(N)$\times$SU$_{-}$(N)
gauge symmetries. However, they do explicitly break the
BRST supersymmetries, and in particular the $\Psi$
independence in the topological sectors, but in a controllable
fashion.
Since the BRST transformation is a change of variables
of the form
\[
\phi^a \ \to \ \phi^a \ + \ \delta \Psi \{ \Omega ,\phi^a \}
\]
we conclude that when we vary the gauge fermion
$\Psi \to \Psi + \delta \Psi$ the phase 
factors do not remain intact but suffer a nontrivial change of
variables. Besides the terms that we have 
discussed in the previous sections, we should then add 
the phase factors which have been subjected to the
proper changes of variables. These are additional
non-local terms that should be included to our action. However,
since these terms are additive and highly complicated, 
we can use general arguments to conjecture that they 
become irrelevant
operators in the infrared limit.

\vskip 0.3cm

\section{Conclusions}

\vskip 0.2cm

In conclusion, by applying
background field formalism in the path integral 
approach we have introduced new variables
to describe ordinary Yang-Mills theory. 
Using our new variables, we 
have then established that in the instanton
approximation ordinary Yang-Mills theory {\it
contains} the N=2 supersymmetric Yang-Mills theory.
Furthermore, by considering all possible solutions to the
Yang-Mills equations, we have established
that this N=2 supersymmetry is embedded in the
Parisi-Sourlas supersymmetry. Finally, we
have found that the
SU(N) gauge symmetry of our Yang-Mills theory
becomes extended into an SU(N)$\times$SU(N) 
gauge symmetry. Such 
extensions of the original 
Yang-Mills symmetry
opens the interesting possibility to explicitly
break the extra symmetries back into the
original SU(N) gauge invariance.
We have investigated in detail a
particular explicit breaking of the additional
symmetries, with intriguing physical implications.
Our symmetry breaking emerges, when 
we add a mass term to the field that describes 
quantum fluctuations around the classical solutions 
in the background field formalism. We have argued 
that this mass term is consistent both with power 
counting renormalizability and unitarity. In 
this way we then obtain an
extension of the original Yang-Mills theory, 
with a gauge invariant mass scale. We have argued that
our construction suggests an approach to color confinement by
the Parisi-Sourlas dimensional reduction mechanism.
In this picture, the large distance fluctuations of
the Yang-Mills vacuum can be viewed as a medium
of randomly distributed
color-electric and color-magnetic fields. We have
derived the relevant  equations of motion 
from a first principle. Since the Parisi-Sourlas
extended Yang-Mills theory 
contains the N=2 supersymmetric Yang-Mills theory in its
selfdual and anti-selfdual sectors, we have also concluded
that this approach to confinement in ordinary Yang-Mills
theory coincides with
the recent proposal to describe confinement in N=2
supersymmetric theories. In particular, the area law
for Wilson loops derived in the Parisi-Sourlas theory
should immediately extend to the N=2 supersymmetric case.

Finally, we point out that 
recently Polyakov \cite{polyakov} has
proposed a string version of Yang-Mills theory, 
that also exhibits Parisi-Sourlas supersymmetry.
We view this as a further indication, that
the Parisi-Sourlas dimensional reduction is indeed the
underlying reason for color and quark confinement
in ordinary Yang-Mills theory. Indeed,
the 4+2 dimensional Parisi-Sourlas supersymmetric
Yang-Mills theory is
intimately connected with the two dimensional 
ordinary Yang-Mills theory and the latter admits
a natural string interpretation \cite{qcdstring}. This
suggests, that there should also be an intimate
relationship between the string theory constructed
by Polyakov and the string theory that describes the
two dimensional Yang-Mills theory. The understanding
of this relationship might
provide an important clue for constructing the string 
variables of four dimensional
Yang-Mills theory. 

\vskip 1.5cm
J.K. thanks O. Tirkkonen and A.N. thanks L. Faddeev, 
A. Polyakov, G. Semenoff, 
N. Weiss and A. Zhitnitsky for discussions. 
We both thank G. Semenoff and
the Department of Physics at University of British
Columbia for hospitality during this work.


\vskip 1.5cm

\begin{thebibliography}{9}

\bibitem{nati} N. Seiberg and E.Witten, Nucl. Phys.
{\bf B431} (1994) 484; {\it ibid.} {\bf
B426} (94) 19

\bibitem{thooft} G. 't Hooft, Nucl. Phys. {\bf B190}
(1981) 455; {\it ibid.} {\bf B153} (1979) 141; A. 
Polyakov, Nucl. Phys. {\bf B120} (1977) 429

\bibitem{bff} J. Honerkamp, Nucl. Phys. {\bf 
B48} (1972) 269

\bibitem{witten} E. Witten, Comm. Math. 
Phys. {\bf 117} (1988) 353

\bibitem{parisi} G. Parisi and N. Sourlas, Phys.
Rev. Lett. {\bf 43} (1979) 744; B. McClain,
A.J. Niemi, C. Taylor and L.C.R. Wijewardhana,
Nucl. Phys. {\bf B217} (1983) 430; Ann. Phys.
{\bf 140} (1982) 232

\bibitem{kopis} H.B. Nielsen and P. Olesen,
preprint NBI-HE-79-45 (unpublished); P. Olesen, 
Nucl. Phys. {\bf B200} (1981) 82; P. Olesen,
Phys. Scripta {\bf 23} (1981) 1000; G. Parisi,
in {\it High Energy Physics - 1980: Proceedings}
(L. Durand and L.G. Pondrom, eds.) (AIP 1981)

\bibitem{kupi} A. Kupiainen and A.J. Niemi,
Phys. Lett. {\bf 120B} (1983) 399

\bibitem{maillet} J.M. Maillet and A.J. Niemi,
Phys. Lett. {\bf B223} (1989) 195; A. Hietam\"aki,
A.J. Niemi and O. Tirkkonen, Phys. Lett. {\bf B238}
(1990) 291

\bibitem{morse} J. Milnor, {\it Morse Theory} 
(Princeton University
Press, 1973)

\bibitem{wit2}  E. Witten, Nucl. Phys. {\bf 
B202} (1982) 253

\bibitem{blau} D. Birmingham, M. Blau, M. 
Rakowski and G. Thompson,
Phys. Rept. {\bf 209} (1991) 129

\bibitem{palo} A.J. Niemi and K. Palo,
hep-th/9406068

\bibitem{dh} M. Blau, E. Keski-Vakkuri and 
A.J. Niemi, Phys. Lett. {\bf B246} (1990) 92; 
A.J. Niemi and O. Tirkkonen, Phys. Lett. {\bf 
B293} (1992) 339

\bibitem{bgv} N. Berline, E. Getzler and M. 
Vergne, {\it Heat Kernels and Dirac
Operators} (Springer Verlag, Berlin, 1991)

\bibitem{atje} M.F. Atiyah and L. Jeffrey,
J. Geom. Phys. {\bf 7} (1990) 120

\bibitem{osv} S. Ouvry, R. Stora and P. 
van Baal, Phys. Lett.
{\bf B220} (1989) 1590

\bibitem{henne}I.A. Batalin and E.S. Fradkin, 
Phys. Lett.
{\bf 122B} (1983) 157; M. Henneaux,
Phys. Rept. {\bf 128} (1986) 1

\bibitem{llew} C.H. Llewellyn Smith, 
Phys. Lett. {\bf 46B}
(1973) 233; J.M. Cornwall, D.N. Levin 
and G. Tiktopoulos,
Phys. Rev. Lett. {\bf 30} (1973) 1268; 
Phys. Rev. {\bf D10}
(1974) 1145

\bibitem{omat} A.J. Niemi, hep-th/9502055 

\bibitem{polyakov} A. Polyakov, seminar at
the University of British Columbia, January 1996

\bibitem{qcdstring} E. Abdalla and M.C.B. Abdalla,
Phys. Rept. {\bf 265} (1996) 253

\end{thebibliography}
\end{document}